\documentclass[english]{iopart}

\usepackage{graphicx}
\usepackage{url}
\usepackage{iopams} 

\expandafter\let\csname equation*\endcsname\relax
\expandafter\let\csname endequation*\endcsname\relax

\usepackage{amssymb}
\usepackage[latin1]{inputenc}
\usepackage{amsmath}
\usepackage{epsfig}
\newcounter{fig}
\begin{document}

\title[Diagonals of rational functions]
{\Large Diagonals of rational functions and selected differential Galois groups}

\vskip .3cm 

\author{A. Bostan$^\P$,  S. Boukraa$||$, 
J-M. Maillard$^\dag$, J-A. Weil$^\pounds$}

\address{$^\P$ \ INRIA, B\^atiment Alan Turing, 1 rue Honor\'e d'Estienne d'Orves,
 Campus de l'\'Ecole Polytechnique, 91120 Palaiseau, France}

\address{$||$  \ LPTHIRM and IAESB,
 Universit\'e de Blida, Algeria}

\address{$^\dag$ LPTMC, UMR 7600 CNRS, 
Universit\'e de Paris 6, Sorbonne Universit\'es, 
Tour 23, 5\`eme \'etage, case 121, 
 4 Place Jussieu, 75252 Paris Cedex 05, France} 

\address{$^\pounds$ \  XLIM, Universit\'e de Limoges, 
123 avenue Albert Thomas,
87060 Limoges Cedex, France}

\vskip .2cm 

{\em Dedicated to R.J. Baxter, for his 75th birthday.}

\begin{abstract}

We recall that diagonals of rational functions naturally occur 
in lattice statistical mechanics and enumerative combinatorics. 
In all the examples emerging from physics,  the minimal 
linear differential operators annihilating these diagonals of rational 
functions have been shown to actually possess
orthogonal or symplectic differential Galois groups. In order to
understand the emergence of such orthogonal or symplectic groups,
we analyze exhaustively three sets of diagonals of rational functions,
corresponding respectively to rational functions of three variables, 
four variables and six variables. We impose the constraints that
the degree of the denominators in each variable is at most one, 
and the coefficients of the monomials are $\, 0$ or $\, \pm 1$,
so that the analysis can be exhaustive.  
We  find the minimal linear differential operators annihilating 
the diagonals of these rational functions of three, four, 
five and six variables. We find that, even for these 
sets of examples which, at first sight,  have no relation with physics, 
their differential Galois groups are always orthogonal or symplectic 
groups. We discuss the conditions on the rational 
functions such that the operators annihilating their diagonals
do not correspond to orthogonal or symplectic differential Galois 
groups, but rather to generic special linear groups.

\end{abstract}

\vskip .1cm

\vskip .4cm

\noindent {\bf PACS}: 05.50.+q, 05.10.-a, 02.30.Hq, 02.30.Gp, 02.40.Xx

\noindent {\bf AMS Classification scheme numbers}: 34M55, 
47E05, 81Qxx, 32G34, 34Lxx, 34Mxx, 14Kxx 

\vskip .2cm

{\bf Key-words}: Diagonals of rational functions, 
differential Galois groups, Hadamard products, 
series with integer coefficients, globally bounded series, 
differential equations Derived From Geometry, 
 modular forms, modular curves, lattice Green functions, 
orthogonal and symplectic groups. 

\vskip .1cm

\section{Introduction}
\label{Introduction}

In  previous papers~\cite{Short,Big} it has been shown
 that the $\, n$-fold integrals $\, \chi^{(n)}$ corresponding 
to the $\, n$-particle contributions
of the magnetic susceptibility of the Ising 
model~\cite{ze-bo-ha-ma-05c,High,Khi6}, as well as various other 
$\,n$-fold integrals of the ``Ising class''~\cite{bo-ha-ma-ze-07b,CalabiYauIsing}, 
or $\, n$-fold integrals from enumerative combinatorics, like 
lattice Green functions~\cite{GoodGuttmann,GlasserGuttmann,DiffAlgGreen,LGF}, 
correspond to a distinguished class of functions {\em generalizing 
algebraic functions}: they are actually 
{\em diagonals of rational functions}~\cite{Purdue,Lipshitz,Christol84,Christol85,Christol369}.
As a consequence, the power series
expansions of the analytic solutions at $\, x=0$ 
of these linear differential equations 
are ``Derived From Geometry''~\cite{bo-bo-ha-ma-we-ze-09} 
and are {\em globally bounded}~\cite{Christol}, 
which means that, after a rescaling of the expansion variable, they can 
be cast into series expansions with {\em integer coefficients}~\cite{Short,Big}. 

In another paper~\cite{DiffAlgGreen} revisiting miscellaneous linear differential 
operators mostly associated with lattice
Green functions in arbitrary dimensions, but also Calabi-Yau operators and 
order-seven linear differential operators corresponding 
to exceptional differential Galois groups, 
we showed that these irreducible operators
are not only {\em globally nilpotent}~\cite{bo-bo-ha-ma-we-ze-09}, but are 
also {\em homomorphic to their} (formal) {\em adjoints}. Considering 
these linear differential operators, or, sometimes, equivalent
operators\footnote[1]{In the sense of the  equivalence of linear differential 
operators, see~\cite{vdP}.}, we showed that, either their 
symmetric square or their exterior square, 
{\em have a rational solution}~\cite{DiffAlgGreen}.
This is a general result: an irreducible\footnote[8]{Minimal order operators
annihilating diagonals of rational functions may be irreducible, in general. 
In that case, all their irreducible factors have the above property.}
linear differential operator
homomorphic to its (formal) adjoint is necessarily such that either its 
{\em symmetric square}, or its {\em exterior square} has a {\em rational solution},
and this situation corresponds to the occurrence of a special differential 
Galois group~\cite{DiffAlgGreen}. We thus defined the notion of 
being ``Special Geometry'' for a linear differential operator
if it is irreducible, globally nilpotent, and such that it is homomorphic 
to its (formal) adjoint~\cite{SG}.  
Since many ``Derived From Geometry''~\cite{bo-bo-ha-ma-we-ze-09} $\, n$-fold 
integrals (``Periods'') occurring in physics are seen 
to be {\em diagonals of rational functions}~\cite{Short,Big}, we address
several examples of (minimal order) operators annihilating
diagonals of rational functions, and remark that they also seem to be,
 almost systematically, associated with irreducible factors 
{\em homomorphic to their adjoints}~\cite{DiffAlgGreen}.

Finally in a last paper~\cite{Canonical} revisiting
 an order-six linear differential operator, already
introduced in~\cite{DiffAlgGreen}, having a solution
which is the diagonal of a rational function of three variables,
we saw that the corresponding
linear differential operator is such that its exterior square has 
a rational solution, indicating that it has a selected 
differential Galois group, and that it is actually 
{\em homomorphic to its adjoint}. We obtained the two 
corresponding intertwiners giving 
this homomorphism to the adjoint relation. We showed that these 
intertwiners are also homomorphic to their adjoints~\cite{Canonical}
 and have a simple decomposition, already underlined in a previous 
paper~\cite{DiffAlgGreen}, in terms of order-two self-adjoint 
operators. From these results,
we deduced a new form of decomposition of operators for this 
selected order-six linear differential operator in terms 
of three order-two self-adjoint 
operators. We generalized this decomposition to decompositions in terms 
of arbitrary self-adjoint operators of arbitrary orders, provided the 
orders have the same parity~\cite{Canonical}. 

This natural emergence in physics of $\,n$-fold integrals that are
diagonals of rational functions, such that their associated linear 
differential operators correspond to selected differential Galois groups,
$\, SO(n, \, \mathbb{C})$ or $\, Sp(n, \, \mathbb{C})$ 
(or subgroups, like exceptional
groups~\cite{DiffAlgGreen}), was illustrated on important problems of
lattice statistical mechanics like the $\,n$-fold integrals 
$\, \chi^{(5)}$ and $\, \chi^{(6)}$ of the square Ising model~\cite{SG},
or non-trivial lattice Green functions examples~\cite{LGF}.

The occurrence of diagonals of rational functions necessarily 
yields linear differential operators that are~\cite{bo-bo-ha-ma-we-ze-09} 
{\em  globally nilpotent}\footnote[9]{And, thus, rational number exponents, for all
the singularities of these Fuchsian equations, wronskians of these
linear differential operators that are $\, N$-th roots 
of rational functions, etc ...}, with series-solutions that are 
{\em globally bounded}\footnote[8]{They can be recast into series with integer
coefficients~\cite{Short,Big}.}~\cite{Christol}, and such that these series
identify {\em modulo prime}, or even 
{\em powers of primes}\footnote[3]{And, in fact, 
modulo any integer.} to  series expansions of 
{\em algebraic functions}~\cite{Short,Big}. We may call these
 (generally transcendental)
holonomic functions ``quasi-algebraic'' transcendental functions. 

If the natural emergence in physics of diagonals of rational functions
also corresponds to highly selected linear differential operators,
the fact, that we observe on all our (large number of quite non-trivial) 
examples of lattice statistical 
mechanics~\cite{High,Khi6,SG,ze-bo-ha-ma-04,mccoy3,ze-bo-ha-ma-05b,CalabiYauIsing1}, 
enumerative combinatorics~\cite{DiffAlgGreen,LGF},  
that such selected linear differential operators
have {\em systematically}  {\em special differential Galois groups},  
is {\em not well understood}. The occurrence of selected differential 
Galois groups ($SO(n, \, \mathbb{C})$, $\, Sp(n, \, \mathbb{C})$, 
or subgroups of $SO(n, \, \mathbb{C})$, 
$\, Sp(n, \, \mathbb{C})$, like exceptional groups such as $\, G_2$ see~\cite{Canonical}) 
is now understood, after~\cite{DiffAlgGreen}, as the fact that the corresponding 
linear differential operators are {\em non-trivially homomorphic to their adjoints}. 
However the ``quasi-algebraic'' character of the diagonal of rational functions 
(reduction to algebraic functions modulo primes or powers of primes)
is {\em not enough} to yield the property, for the corresponding linear differential 
operator, to be homomorphic to its adjoint.

We have accumulated a very large number of examples\footnote[4]{Unpublished results.} 
of diagonals of rational functions for which, quite systematically, the 
corresponding linear differential operators are homomorphic to their adjoints, 
thus having selected differential Galois groups. The set of diagonals of 
rational functions is an extremely large set: an accumulation
of examples, that one can hope to be representative of the generic case,
is just a way to build educated guess, or intuition. 
Among such results the exhaustive analysis of certain sets is worth mentioning.
For instance, a set of 210  explicit linear differential
operators annihilating periods,
which are {\em actually diagonals of rational
functions}, arising from {\em mirror symmetries}\footnote[2]{Using 
a smoothing criterion of Namikawa~\cite{Namikawa}, Batyrev and
 Kreuzer found~\cite{Kreuzer} 30241 
reflexive 4-polytopes such that the corresponding Calabi-Yau hypersurfaces 
are smoothable by a flat deformation. In particular, they 
found 210 reflexive 4-polytopes 
defining 68 topologically different Calabi-Yau 3-folds with $\, h_{11}= \, 1$.}
 (associated with reflexive 4-polytopes 
defining 68 topologically different 
Calabi-Yau 3-folds, see~\cite{Kreuzer,Lairez}) obtained
 by P.~Lairez~\cite{Lairez}, has been analyzed in~\cite{Canonical}.
Among these 210 operators many correspond to ``standard'' 
Calabi-Yau ODEs, already analyzed in various 
papers~\cite{CalabiYauIsing}\footnote[1]{They are {\em order-four} 
irreducible operators satisfying  the ``{\em Calabi-Yau condition}: they are, up 
to a conjugation by a function, irreducible {\em order-four self-adjoint} 
operators~\cite{DiffAlgGreen}. This amounts to saying that the exterior 
square of these order-four operators is of {\em order five}.}. However, 
remarkably, the other linear differential operators 
are (non classical Calabi-Yau) {\em higher order}
operators of even orders $\, N \, = \, \, 6, \, 8, \, 10,$
$  \, \cdots ,  \, 24$. It was found that all these 
linear differential operators have {\em symplectic} differential 
Galois groups, $\, Sp(N, \, \mathbb{C})$, with a remarkable canonical 
decomposition~\cite{Canonical}, in terms of self-adjoint order-two  operators, 
except the self-adjoint ``rightmost operator'' which is
 systematically of {\em order four}.

After these accumulations of examples of diagonals of rational functions yielding 
systematically selected differential Galois groups, one might be led to conjecture
that the  linear differential operators annihilating diagonals of rational 
functions are necessarily 
homomorphic to their adjoints. Such a conjecture is in fact trivially false, 
as can be seen with the simple $\, _3F_2$ hypergeometric example 
$\,_3F_2([1/3, \, 1/3, \, 1/3], \, [1, \, 1], \,  3^6 \, x)$ 
which is the {\em Hadamard product}~\cite{Short,Big,Hadamard,CaMa38} of 
three times the simple algebraic function 
$\, (1\, -3^2\, x)^{-1/3}$ (Hadamard cube), and is, thus, 
the diagonal of a rational function~\cite{Short,Big},
the corresponding order-three linear differential 
operator having a $\, SL(3, \, \mathbb{C})$  differential Galois group.
In other words this operator, associated with the diagonal of a 
rational function, {\em cannot be homomorphic to its adjoint} 
(even with an algebraic extension). One can find many other similar 
counter-examples of diagonals of a rational function with a 
$\, SL(3, \, \mathbb{C})$  differential Galois group.
For instance, 
$\,_3F_2([1/3, \, 1/3, \, 1/5], \, [1, \, 1], \,  3^4 \, 5^2 \, x)$
which is the {\em Hadamard product} of two times $\, (1\, -3^2\, x)^{-1/3}$
with $\, (1\, -5^2\, x)^{-1/5}$, or 
$\,_3F_2([1/2, \, 1/2, \, 1/3], \, [1, \, 1], \,  2^4 \, 3^2 \, x)$
which is the {\em Hadamard product} of two times  the
 algebraic function $\, (1\, -2^2\, x)^{-1/2}$
with $\, (1\, -3^2\, x)^{-1/3}$, are two such examples.

The property, for a linear differential operator, to be 
homomorphic to its adjoint, can be seen to be ``some kind''\footnote[9]{To
be more explicit would require to write a mathematical paper, calling out
a Deligne-Steenbrink-Zucker theorem saying that Gauss-Manin connections
are ``variations of polarized mixed Hodge structures'', the associated
graded modules being Gauss-Manin connections of {\em smooth} projective varieties,
and that they are {\em self-adjoint by Poincar\'e duality}. This is far beyond 
the scope of this ``learn-by-example'' paper (see for instance~\cite{Sabbah}).} of 
{\em Poincar\'e duality}~\cite{Griffiths,Sabbah}. Actually, considering a diagonal 
of a rational function, amounts to considering an algebraic 
variety. If this algebraic variety is ``smooth enough'' (not too singular), 
then one will have a Poincar\'e duality\footnote[1]{The Poincar\'e
 duality theorem is a result on the structure of the 
homology and cohomology groups of manifolds 
(see chapter 4, page 53 of~\cite{Griffiths}).},
which, in some abstract $\, \cal{D}$-module perspective~\cite{Coutinho}, should 
correspond to the previous ``homomorphism to the adjoint'' property. 
A simpler occurence of this phenomenon is shown by Bogner in proposition 3.4,
 page 5 of~\cite{Bogner}. In simpler words, the emergence of a 
selected differential Galois group should be
natural for the diagonals of a rational function, when the rational function
is not ``too singular''. 

For diagonals of rational functions that we know to occur naturally in 
theoretical physics\footnote[2]{And we know why, see~\cite{Short,Big}.}, 
understanding the emergence of such a duality yielding selected differential 
Galois groups, at least in a physicist's perspective\footnote[5]{Who wants to 
understand why orthogonal, symplectic or exceptional groups occur in 
his problems.} 
requires the analysis of other well-defined sets of diagonals of rational functions, 
along the line of the previously mentioned exhaustive analysis~\cite{Short,Big}
 of the 210  explicit linear differential operators arising from 
mirror symmetries.

This paper will provide such an analysis, introducing well-defined sets of
diagonals of rational functions of three, four and six variables, 
showing that all these examples yield 
selected differential Galois groups, namely {\em orthogonal or symplectic} groups.
We will first analyze a large set of rational functions of {\em three}  variables, 
namely rational functions with 
denominators with degree bounded by $1$. It turns out that all these diagonals  
of rational functions of {\em three}  variables
correspond to {\em modular forms}, so that  all can, in principle, 
be written as $\, _2F_1$ hypergeometric functions with 
{\em two pull-backs}~\cite{IsingModularForms}.

We will finally provide miscellaneous examples that are worth keeping in mind
when one tries to understand in which case selected differential Galois groups
do not occur for diagonals of rational functions, so that one could see if such
``exceptional'' cases can also occur in a physics framework. 

\vskip .2cm 

\section{Diagonals of rational functions: preliminary comments }
\label{diag}

Let us recall the definition of the diagonal of a rational function of $\, n$ variables
$\,{\cal R}(x_1, \ldots, x_n)\, = \,\, {\cal P}(x_1, \ldots, x_n)/{\cal Q}(x_1, \ldots, x_n)$ 
 where $ {\cal P}$ and $ {\cal Q}$ are polynomials of $\, x_1, \, \cdots, \, x_n\,$ 
with {\em rational coefficients} such that $\, {\cal Q}(0, \ldots, 0) \neq 0$. The 
diagonal of ${\cal R}$ is defined 
from its multi-Taylor expansion
\begin{eqnarray}
\label{defdiag}
\hspace{-0.90in}&&\quad \quad \, \, \,
{\cal R}\Bigl(x_1, \, x_2, \, \ldots, \, x_n \Bigr)
\, \, \,\, = \, \,\, \,\sum_{m_1 \, = \, 0}^{\infty}
 \, \cdots \, \sum_{m_n\, = \, 0}^{\infty} 
 \,R_{m_1,  \, \ldots, \, m_n}
\cdot  \, x_1^{m_1} \,\,  \cdots \,\, x_n^{m_n}, 
\end{eqnarray}
as the series
of {\em one variable}
\begin{eqnarray}
\label{defdiag2}
\hspace{-0.7in}&&\quad \quad Diag\Bigl({\cal R}\Bigl(x_1, \, x_2, \, \ldots, \, x_n \Bigr)\Bigr)
\, \, \, = \, \,  \quad \sum_{m \, = \, 0}^{\infty}
 \,R_{m, \, m, \, \ldots, \, m} \cdot \, x^{m}.
\end{eqnarray}

In the generic case, denoting $\, d$ the degree of 
the denominator of a rational function of $\, n$ variables, and  $\, \Omega$
the order of the minimal order linear differential operator annihilating 
the diagonal of this rational function, the order $\, \Omega$  grows 
with $\, n$ and $\, d$ like $\,\simeq  d^n$ (see e.g.~\cite{Lairez3}):
 \begin{eqnarray}
\label{formula}
\hspace{-0.95in}&& \quad \quad \quad \quad \quad \,
\Omega \, \,\, = \, \, \, \,\,
{{1} \over {d}} \cdot \, 
\Bigl((d-1)^{n+1} \, + \, (-1)^{n+1} \cdot \,  (d-1)\Bigr) 
\,\,\, \, <\,\, \,\, d^n.
\end{eqnarray} 
From a {\em theory of singularity} perspective~\cite{Hironaka}, one 
can imagine that the analysis of such diagonals of  
rational functions depends ``essentially''
 on the denominator of the rational function.
For that reason we will restrict most of the examples 
of this paper to rational functions with a numerator
normalised to $\, 1$, {\em so that the number of cases
to analyze will be reasonable}. Note, however, that,
even in a theory of singularity perspective, this restriction 
to numerators equal to a constant, is not innocent, as 
will be seen below (see the Remark in section (\ref{nonhomosub})).
We impose this restriction, for the simplicity 
of the calculations, and in order to perform
exhaustive analysis of certain sets of examples. 

\vskip .1cm 

Furthermore, in order to be able to 
find  a linear differential operator for the diagonal, using reasonable 
computer resources, we will, for a given number of variables $\, n$,
restrict to denominators of the lowest possible degree, imposing, 
for instance, that the degree, in each variable $\, x_i$, 
of the monomials $\, x_1^{d_1} \, x_2^{d_2} \,  \cdots \, x_n^{d_n}$
of the denominator of the rational function is {\em at most $\, 1$}.  
Formula (\ref{formula}) gives the upper bound for the order of the linear 
differential operator, which is actually reached when the polynomial 
at the denominator, has {\em all its monomials} (no monomial
has a zero coefficient). In practice, in the denominator of the rational 
function examples emerging from physics, the polynomial at the denominator is 
(fortunately) sparse, being the sum of a {\em quite small 
set of monomials}. In such (physical) cases, the order of the corresponding 
linear differential operator is less than the one given by (\ref{formula})
which grows like $\,\simeq \,  d^n$ and thus  becomes quickly too large
for any formal calculation.   

Imposing a constraint on the degree in each variable, instead of a 
constraint on the degree of the denominator $\, d \, = $
$\, \,d_1  + d_2\, + \, \cdots \, +  d_n$, reduces the number of 
monomials and, consequently, reduces quite drastically, the order 
of the corresponding linear differential operator. 

Furthermore, it is clear (by definition of the diagonal of a function)
that scaling the variables 
$\, x_i \, \rightarrow \, \, \lambda_i \cdot \, x_i$,
amounts to performing a simple scaling on the 
diagonal: $\, \Delta(x) \,  \rightarrow \, \, $
$\Delta(\lambda \cdot \, x)$  where $\, \lambda \, =\, \, $
$\lambda_1 \cdot \, \lambda_2 \, \cdots  \lambda_n$. 
Consequently we will, in this paper, often restrict the coefficients of 
the monomials to a narrow set of {\em small} integer values, namely
 $\, 0, \, 1$, or  $\, 0, \, \pm 1$, 
in order to have a reasonably small number of cases to analyze
to avoid any explosion of the combinatorics. It is, of course, clear that 
the coefficients of  rational functions cannot, in general, be reduced to 
 $\, 0, \, \pm 1$.

Diagonals of rational functions of two variables necessarily yield {\em algebraic 
functions}~\cite{Short,Big,Fu,Denef,BA-JPB,Polya21}. Therefore, 
in the next sections, we will consider diagonal of 
rational functions, of {\em more than two variables}.

\vskip .1cm

\section{Diagonals of rational functions of three variables }
\label{three}

Let us, first, consider diagonals of rational functions  
of the form $\, 1/(1 \, -P(x, \, y, \, z))$ where $\, P(x, \, y, \, z)$
is a polynomial of {\em three} variables $\, x, \, y, \, z$, 
sum of monomials $\, x^m \, y^n \, z^p$ where the degrees $\, m, \, n, \, p$
are $\, 0$ or $\, 1$, and where the coefficients in front of these 
monomials are restricted to take two values $\, 0, \,  1$. We will say
that two rational functions of the form $\, 1/(1\, -P)$ are in the same class
if they
have the same diagonal and hence the same linear differential operator
annihilating this diagonal.
With these two constraints (on the degrees and values of the coefficients),
one finds only $\, 20$ classes of rational functions (or diagonals). For 
all these cases, the linear differential operator is of order two, 
the diagonal being a $\, _2F_1$ hypergeometric function,
or a HeunG function~\cite{IsingModularForms} that can be 
rewritten as a $\, _2F_1$ hypergeometric 
function with {\em two possible pullbacks}, which means that this is, 
in fact, a {\em modular form}~\cite{IsingModularForms}. 

\vskip .1cm 

With the same constraint on the degrees $\, m, \, n, \, p$
to be $\, 0$ or $\, 1$, restricting  the coefficients in front of these 
monomials  to take three values $\, 0, \, \pm 1$, one obtains $ 85$ 
classes of linear differential operators.

\vskip .1cm 

\vskip .1cm 

Let us just give a few examples, the exhaustive list of 
results being given in a web page of supplementary 
material~\cite{supplementary}.

\vskip .1cm 

$\bullet$  For the polynomial $\, P(x, \, y, \, z) \, = \,  \,$
$x  \,+ y  \,+ z \, - x \, z \, - y \, z \, +x\, y\, z$, 
the diagonal of $\, 1/(1 \, -P(x, \, y, \, z))$, which corresponds 
to the sequence $\,[1,\, 3,\, 13, \,63, \,321,\, 1683, \, \cdots ]$ 
of Central Delannoy numbers (see Sloane's 
on-line encyclopedia~\cite{Sloane} of integer sequences: A001850),
is a simple {\em algebraic function} 
$\, (1 \,-6 \,x \,+ \,x^2)^{-1/4}$, annihilated by an order-one operator.

\vskip .1cm 

$\bullet$ For the polynomial $\, P(x, \, y, \, z) \, = \,  \,$
$x  \,+ y  \,+ z \, - x \, z \, -x\, y\, z$, 
the diagonal of $\, 1/(1 \, -P(x, \, y, \, z)) \, $ is the 
pullbacked $\, _2F_1$ hypergeometric function
\begin{eqnarray}
\label{A005258}
 \hspace{-0.95in}&& \quad \quad     
(1 \, -12\,x \, +14\,{x}^{2} \, +12\,{x}^{3} \, +{x}^{4})^{-1/4} \cdot \, 
 \,_2F_1\Bigl([{{1} \over {12}}, \, {{5} \over {12}}], \, [1], \, P_1(x)\Bigr) 
\nonumber \\
\hspace{-0.95in}&& \quad \quad  \quad  \quad     \hbox{where:} 
\quad\quad \quad   \quad  
P_1(x)\, \,\, = \, \, \,  \, \, 
1728\,{\frac {{x}^{5} \cdot \, (1\, -11\,x\, -{x}^{2}) }{ 
(1 \, -12\,x \, +14\,{x}^{2} \, +12\,{x}^{3} \,+ {x}^{4})^{3}}}.
\end{eqnarray}
This {\em Hauptmodul}~\cite{IsingModularForms} $\, P_1(x)$ 
corresponds to the Hauptmodul $\, 12^3/j_5'$
(see Table 5 in~\cite{Maier1}) 
\begin{eqnarray}
\label{j5}
 \hspace{-0.95in}&& \quad \, \,    \quad  
P_1(x) \, \, = \, \, \,
 {{ 12^3 \cdot \, z^5} \over {(z^2 \, +250\,z\,+3125)^3 }} 
\qquad \hbox{with:} 
 \qquad  z \, \, = \, \, \,  {{ 5^3 \cdot \, x } \over {1\, -11\,x\, -{x}^{2} }}. 
\end{eqnarray}

This diagonal corresponds to the sequence 
$\, [1, 3, 19, 147, 1251, 11253, \cdots]$  of {\em Apery numbers} (see Sloane's 
on-line encyclopedia~\cite{Sloane} of integer sequences: A005258). 
The pullbacked $\, _2F_1$ hypergeometric function (\ref{A005258})
can also be written {\em with another pullback}
\begin{eqnarray}
\label{A005258bis}
 \hspace{-0.95in}&& \,      \quad     
(1\, +228\,x\,+494\,{x}^{2}\,-228\,{x}^{3}\,+{x}^{4})^{-1/4} \cdot \, 
 \,_2F_1\Bigl([{{1} \over {12}}, \, {{5} \over {12}}], \, [1], \, P_2(x)\Bigr) 
\nonumber  \\
\hspace{-0.95in}&& \quad \quad  \quad     
  \hbox{where:} 
\quad\quad \quad   
P_2(x)\, \,\, = \, \, \,  \, \, 
1728\,{\frac {{x} \cdot \, (1\, -11\,x\, -{x}^{2})^5 }{ 
(1\, +228\,x\,+494\,{x}^{2}\,-228\,{x}^{3}\,+{x}^{4})^{3}}}. 
\end{eqnarray}
This Hauptmodul~\cite{IsingModularForms} $\, P_2(x)$ 
corresponds to the Hauptmodul $\, 12^3/j_5$
(see Table 4 in~\cite{Maier1})
\begin{eqnarray}
\label{j5}
 \hspace{-0.95in}&& \quad \quad    \quad  
P_2(x) \, \, = \, \, \,
 {{ 12^3 \cdot \, z} \over {(z^2 \, +10\,z\,+5)^3 }} \qquad \hbox{with:} 
 \qquad  z \, \, = \, \, \,  {{ 5^3 \cdot \, x } \over {1\, -11\,x\, -{x}^{2} }}. 
\end{eqnarray}
Changing  $\, P_1(x)$ into $\, P_2(x)$ amounts to changing 
$\, z \,  \leftrightarrow \, \, \, 5^3/z$.

These two pullbacks, $\, Y \, = \, \, P_1(x)$ and 
$\, Z \,  =  \, \,P_2(x)$, are related 
by an algebraic curve, namely  the (genus zero) {\em modular  curve}:
\begin{eqnarray}
\label{modularA005258bis}
 \hspace{-0.95in}&& 
 2^{54} \cdot \, 5^3 \cdot \, 11^9 \cdot \, {Y}^{6} Z^6 \, \, 
+ 110949\cdot \, 2^{47} \cdot \, 11^6  \cdot \, 5^3 \,   \cdot  \,{Y}^{5} Z^5 \cdot \, (Y+Z)
\nonumber \\
\hspace{-0.95in}&&      
  +  3\cdot \, 2^{36}    \cdot  \, 11^3  \cdot  \, {Y}^{4} Z^4 \cdot \, 
(2735484611275 \,({Y}^{2}\,+ \, {Z}^{2})\,  \, -107937074856652 \,Y Z)
\nonumber \\
\hspace{-0.95in}&&  
  + 2^{29} \cdot \, 5^2 \cdot \, {Y}^{3} Z^3 \cdot \, (Y+Z) \cdot \, 
(4046657341273198 \,({Y}^{2}\, + \, {Z}^{2}) 
 \nonumber \\
\hspace{-0.95in}&&  \qquad \qquad 
\, +523793662474799327 \cdot \,Y Z)
\nonumber \\
\hspace{-0.95in}&&     
 + 3 \cdot \, 2^{18} \cdot \, 5 \cdot \,  \,{Y}^{2}{Z}^{2} \cdot \, 
(13^2 \cdot \, 647451979 \cdot \, ({Y}^{4}+\,{Z}^{4})
\nonumber \\
\hspace{-0.95in}&&   +5^3 \cdot \, 3482348755357972227 \cdot \,{Y}^{2} {Z}^{2}  
 \, -16391442082714013450 \cdot \, (Y {Z}^{3}\, +{Y}^{3} Z) )
\nonumber \\
\hspace{-0.95in}&& 
 +30720 \cdot \,Y \, Z \cdot \, (Y+Z) \cdot \, 
(36983 \cdot \, ({Y}^{4}\, +\,{Z}^{4})\,\, 
 +2421471845930417 \cdot \, (Y {Z}^{3} \, +{Y}^{3} Z)
 \nonumber \\
\hspace{-0.95in}&& \quad \quad \quad  \quad  \quad 
\, +506711612589929401008 \,{Y}^{2} {Z}^{2})
 \nonumber\\
\hspace{-0.95in}&&
+({Y}^{6}\, +{Z}^{6})\, -246683410950 \cdot \,(Y Z^{5} \,+ Y^{5} Z)\, 
-441206965512914835246100 \cdot \, Y^{3} Z^{3} 
\nonumber \\
\hspace{-0.95in}&&  \quad   \quad   \quad   \quad  
+383083609779811215375 \cdot \,(Y^{4} Z^{2} \,+ \, Z^{4} Y^{2} ) \, 
\nonumber \\
\hspace{-0.95in}&&  \quad  
\, + 2^{11} \cdot \, 3^3 \cdot 5^2  \cdot \,Y\, Z \cdot \, (Y +Z) \cdot \,
 (2535689 \cdot \,({Y}^{2}\, + \,{Z}^{2}) \, \, + 134848657695982 \,Y Z)
\nonumber \\
\hspace{-0.95in}&&  \quad \quad  
\,  -59719680 \cdot \, Y \, Z \cdot \, 
(227547 \cdot \,({Y}^{2}\, + \, {Z}^{2}) \, -83299968230 \cdot \,Y Z)
\nonumber \\
\hspace{-0.95in}&& \quad \quad   \quad  \quad  
\,  +  2^{21} \cdot \, 3^{10} \cdot \, 5  \cdot \,  31  \cdot \, Y \, Z \cdot \, (Y +Z)\, 
\,\, \, - 2^{24} \cdot \, 3^{12}  \cdot\, Y \,Z  
\, \,\,\, = \, \, \,\, \, 0.
\end{eqnarray}
The pullbacked $\, _2F_1$ hypergeometric function  (\ref{A005258}),
or equivalently (\ref{A005258bis}), is a {\em modular form} (see 
the $\, J_5$ of Maier~\cite{Maier1}).

\vskip .1cm 

$\bullet$ For polynomial $\, P(x, \, y, \, z) \, = \,  \,$
$x  \,+ y  \, + x \, z \, + \, y \, z  \,+x\, y\, z$, 
the diagonal of $\, 1/(1 \, -P(x, \, y, \, z))$ is the 
pullbacked $\, _2F_1$ hypergeometric
function
\begin{eqnarray}
\label{A243945}
 \hspace{-0.95in}&& \quad \,   
(1 \, -20\,x \, +54\,{x}^{2} \, -20\,{x}^{3} \, +{x}^{4})^{-1/4} \cdot \, 
 \,_2F_1\Bigl([{{1} \over {12}}, \, {{5} \over {12}}], \, [1], \, P_1(x)\Bigr) 
\nonumber \\
\hspace{-0.95in}&& \quad  \quad  \quad   \quad  \quad   
    \hbox{where:} 
\quad\quad \quad     
P_1(x)\, \,\, = \, \, \,  \, \, 
1728\,{\frac {{x}^{4} \cdot \, (1-18\,x \, +x^2)  \, (x-1)^{2}}{ 
(1 \, -20\,x \, +54\,{x}^{2} \, -20\,{x}^{3} \, +{x}^{4})^{3}}}. 
\end{eqnarray}
This Hauptmodul $\, P_1(x)$ corresponds to the Hauptmodul $\, 12^3/j_2'$
(see Table 5 in~\cite{Maier1}) 
\begin{eqnarray}
\label{j5}
 \hspace{-0.95in}&& \quad \,      
P_1(x) \, \, = \, \, \,
 {{ 12^3 \cdot \, z^2} \over {(z \, + \, 256)^3 }} \qquad \hbox{with:} 
 \quad \quad  z \, \, = \, \, \, 
 \,{\frac {  2^{12} \cdot \,  {x}^{2}}{ (1 \,-x)^{2} \, (1 \, -18\,x \, +{x}^{2}) }}. 
\end{eqnarray}

This diagonal corresponds to the sequence 
$\, [1, 5, 49, 605, 8281, 120125 \cdots]$ (see Sloane's on-line encyclopedia 
of integer sequences: A243945). 
Similarly, the  pullbacked $\, _2F_1$ hypergeometric function (\ref{A243945})
can also be written
\begin{eqnarray}
\label{A243945bis}
\hspace{-0.95in}&& \quad   \,   
(1-20\,x+294\,{x}^{2}-20\,{x}^{3}+{x}^{4})^{-1/4} \cdot \, 
 \,_2F_1\Bigl([{{1} \over {12}}, \, {{5} \over {12}}], \, [1], \, P_2(x)\Bigr) 
\nonumber \\
\hspace{-0.95in}&& \quad \quad  \quad  \quad    
   \hbox{where:} 
\quad\quad \quad   
P_2(x)\, \,\, = \, \, \,  \, \, 
1728\,{\frac {{x}^{2} \cdot \, (1-18\,x \, +x^2)^2  \cdot \, (1-x)^{4}}{ 
(1\, -20\,x\, +294\,{x}^{2}\, -20\,{x}^{3}\, +{x}^{4})^{3}}}.
\end{eqnarray}
This Hauptmodul $\, P_2(x)$ corresponds to the Hauptmodul $\, 12^3/j_2$
(see Table 4 in~\cite{Maier1})
\begin{eqnarray}
\label{j2}
 \hspace{-0.95in}&& \quad \,    
P_2(x) \, \, = \, \, \,
 {{ 12^3 \cdot \, z} \over {(z \, +16)^3 }}
\qquad \,  \hbox{with:} \quad \, 
 \quad  z \, \, = \, \, \,
  \,{\frac { 2^{12} \cdot \,  {x}^{2}}{ (1 \,-x)^{2} \, (1 \, -18\,x \, +{x}^{2}) }}. 
\end{eqnarray}
Changing  $\, P_1(x)$ into $\, P_2(x)$ amounts to changing 
$\, z \,  \leftrightarrow \, \, \, 2^{12}/z$. These two pullbacks,
$\, Y \, = \, \, P_1(x)$ and 
$\, Z \,  =  \, \,P_2(x)$, are related 
by an algebraic curve, namely  the (genus zero) {\em modular  curve}:
\begin{eqnarray}
\label{modularA243945bis}
 \hspace{-0.95in}&&  
5^9 \, {Y}^{3} {Z}^{3}\,
 - 5^6 \cdot \, 12\,{Y}^{2}{Z}^{2} \cdot  \, (Y+Z) \, 
 +375 \cdot \,Y\,Z \cdot \, (16\,{Y}^{2} \,
 -4027 \cdot \,Y Z \,+16\,{Z}^{2})
\nonumber \\
\hspace{-0.95in}&& \quad \,   \quad \quad \quad \quad \quad
 -64 \cdot \, (Y +Z)  \cdot \,  ({Y}^{2} +1487 \,Y Z +{Z}^{2})
\,\, +  2^{12} \cdot \, 3^3 \cdot \,Y Z
\,\,\,  = \,\,\,\,  0.    
\end{eqnarray}
The pullbacked $\, _2F_1$ hypergeometric function  (\ref{A243945}),
or equivalently (\ref{A243945bis}), is a {\em modular form} (see 
the $\, J_2$ of Maier~\cite{Maier1}, or even $\, J_4$ of 
Maier~\cite{Maier1} but with $\, z \, = $
$\, \, 2^{12} \, x \cdot \, (1-x)^2/(1 \, -18\,x \, +{x}^{2})^2$).

\vskip .1cm 

$\bullet$ For the polynomial $\, P(x, \, y, \, z) \, = \,  \,$
$x y z + x y + x z + y z + x + y + z$,  
the diagonal of $\, 1/(1 \, -P(x, \, y, \, z))$ is the 
pullbacked\footnote[5]{Note that this pullback can be 
obtained using the Maple program ''hypergeomdeg3'' of 
V. J. Kunwar and M. van Hoeij~\cite{Kunwar,Kunwar2}.} 
$\, _2F_1$ hypergeometric function
\begin{eqnarray}
\label{A126086}
 \hspace{-0.95in}&& \quad \,  {{1} \over {1\, -x}} \cdot \, 
 \,_2F_1\Bigl([{{1} \over {3}}, \, {{2} \over {3}}], \, [1], \, P(x)\Bigr)  
 \quad \quad   \hbox{where:} 
\quad\quad \quad   \,   
P(x)\, \,\, = \, \, \,  \, \, 
\,{\frac { 54 \,  \, x}{ \left( 1-x \right) ^{3}}}.  
\end{eqnarray}
This diagonal corresponds to the sequence 
$\, [1, 13, 409, 16081, 699121, 32193253\, \cdots]$ (see Sloane's 
on-line encyclopedia of integer sequences: A126086).
This series can also be written as the 
 pullbacked $\, _2F_1$ hypergeometric function  
\begin{eqnarray}
\label{alternat}
\hspace{-0.95in}&& \quad    \,    
(1\, -x)^{-1/4} \cdot \, (1 \, +429 \, x \, +3 \, x^2 \, -x^3 )^{-1/4} \cdot \,
 \,_2F_1\Bigl([{{1} \over {12}}, \, {{5} \over {12}}], \, [1], \, P_1(x)\Bigr) 
\nonumber   \\
\hspace{-0.95in}&& \quad \quad    \quad      
      \hbox{where:} 
\quad\quad \quad   \quad    
P_1(x)\, \,\, = \, \, \,  \, \, 
3456\,{\frac {{x} \cdot \, (1 \, -57 \, x \, +3 \, x^2 \, -x^3)^3 }{ 
(1\, -x)^3 \, (1 \, +429 \, x \, +3 \, x^2 \, -x^3)^{3}}}, 
\end{eqnarray}
or as  the 
 pullbacked $\, _2F_1$ hypergeometric function  
\begin{eqnarray}
\label{alternatbis}
\hspace{-0.95in}&& \quad \, \,      
(1\, -x)^{-1/4} \cdot \, (1 \,-51 \,x \,+3 \,x^2 \,-x^3 )^{-1/4} \cdot \,
 \,_2F_1\Bigl([{{1} \over {12}}, \, {{5} \over {12}}], \, [1], \, P_2(x)\Bigr) 
\nonumber   \\
\hspace{-0.95in}&& \quad  \quad  \quad   
      \hbox{where:} 
\quad\quad \quad   \, \,      
P_2(x)\, \,\, = \, \, \,  \, \, 
13824\,{\frac {{x}^3 \cdot \, (1 \, -57 \, x \, +3 \, x^2 \, -x^3)}{ 
(1\, -x)^3 \, (1 \,-51 \,x \,+3 \,x^2 \,-x^3)^{3}}}.  
\end{eqnarray}
These two pullbacks correspond respectively to 
the Hauptmodul~\cite{IsingModularForms} 
 $\, 12^3/j_3$ of Table 4 in~\cite{Maier1}
and $\, 12^3/j_3'$
of Table 5 in~\cite{Maier1}:
\begin{eqnarray}
\label{j3} 
 \hspace{-0.95in}&& \quad \,   \quad
P_1(x) \, \, = \, \, \,
 {{ 12^3 \cdot \, z} \over { (z \, +27)\cdot \, (z \, +3)^3}}, 
 \quad \quad \quad  \quad         
P_2(x) \, \, = \, \, \, 
{{ 12^3 \cdot \, z^3} \over { (z \, +27)\cdot \, (z \, +243)^3}}
\nonumber \\
\hspace{-0.95in}&& \quad \quad  \quad  \quad   \quad  \quad \, \, 
 \qquad \hbox{with:} \quad
 \qquad  z \, \, = \, \, \,
  2 \cdot {{ 9^3 \cdot \, x } \over { 1 \, -57 \,x \, +3 \,x^2 \, -x^3}}. 
\end{eqnarray}
Changing $\, P_1(x)$ into $\, P_2(x)$ amounts to changing 
$\, z \, \leftrightarrow \, 3^6/z$.
Again, these two pullbacks are related by a (genus zero)
{\em modular curve}:
\begin{eqnarray}
\label{modcurve3}
\hspace{-0.95in}&& \quad \,    \,     
 2^{27} \cdot \, 5^9 \cdot \, {Y}^{3} {Z}^{3}\cdot \, (Y +Z)
\, \,  + 2^{18} \cdot \, 5^6  \cdot \,{Y}^{2} {Z}^{2} \cdot \, 
(27\,{Y}^{2} -45946 \,Y Z +27\,{Z}^{2}) \, 
\nonumber \\ 
\hspace{-0.95in}&& \quad \quad   \,   
+ 2^{9} \cdot \, 5^3 \cdot \, 3^5 \cdot \, \,
Y Z \cdot \, (Y +Z)  \cdot \, ({Y}^{2} +241433\,Y Z +{Z}^{2}) 
\nonumber \\ 
\hspace{-0.95in}&& \quad \quad   \,    
\, +729 \cdot  \,({Y}^{4}\, + \,{Z}^{4}) \, \, 
 -779997924 \cdot \, (Y {Z}^{3}) \, +{Y}^{3} Z \, 
 + 31949606 \cdot 3^{10} \cdot \, {Y}^{2} {Z}^{2} 
\nonumber \\ 
\hspace{-0.95in}&& \quad \quad   \,  
+2^9 \cdot \, 3^{11} \cdot \, 31 \cdot \,Y \, Z \cdot \, (Y +Z)
\,  - 2^{12} \cdot \, 3^{12} \cdot \,Y Z\,\,\,= \,\,\,\,0. 
\nonumber   
\end{eqnarray}
The pullbacked $\, _2F_1$ hypergeometric function  (\ref{alternatbis}) 
is a {\em modular form} (see the $\, J_3$ of Maier~\cite{Maier1}).

\vskip .1cm 

$\bullet$  For the polynomial $\, P(x, \, y, \, z) \, = \,  \,$
$x y z + x y + x z + y z$,  
the diagonal of $\, 1/(1 \, -P(x, \, y, \, z))$ is the 
pullbacked $\, _2F_1$ hypergeometric function
\begin{eqnarray}
\label{A126086}
 \hspace{-0.95in}&& \quad   \,  {{1} \over {1\, -x}} \cdot \, 
 \,_2F_1\Bigl([{{1} \over {3}}, \, {{2} \over {3}}], \, [1], \, P(x)\Bigr)  
 \quad \quad   \hbox{where:} 
\quad\quad \,   \quad  
P(x)\, \,\, = \, \, \,  \, \, 
\,{\frac {27 \, \, x^2}{ (1-x)^{3}}}.  
\end{eqnarray}
This diagonal corresponds to the sequence 
$\,[1, 1, 7, 25, 151, 751\, \cdots]$ (see Sloane's on-line encyclopedia 
of integer sequences: A208425).

\vskip .1cm 

In \ref{modular} we give a set of 
diagonals of $\, \, 1/(1 \, -P(x, \, y, \, z))\, $ which can be 
 seen to be {\em modular forms}: they can be written as
 $\, _2F_1$ hypergeometric functions with two different pullbacks.

\vskip .1cm 

Most of the other examples of diagonals, given in a web page as well as 
in the supplementary 
material~\cite{supplementary}, are not yet in Sloane's on-line encyclopedia 
of integer sequences. For instance the sequence 
$\, [1,\,  4, \,  42,\,   520, \,  7090,\,   102144, \, \cdots ]$,
which corresponds to the diagonal of the rational function
$\,\,  1/ (1\,  - x\,  - y \,  -x\,y \, - x\,z\,  - y\,z)$, 
can be written as a $\, _2F_1$ hypergeometric function 
with a rational pullback: 
\begin{eqnarray}
\label{notin1}
 \hspace{-0.95in}&& \, \,\, \,  
\Bigl({{1} \over {1 \, -16 \,x \,-8 \,x^2 }}\Bigr)^{1/4}  
\cdot \, 
 _2F_1\Bigl([{{1} \over {12}}, \, {{5} \over {12}}], \, [1], \, 
1728 \cdot \, 
{{ {x}^{4} \cdot \, (1+x)^{2} (2-34\,x-27\,{x}^{2}) } \over {(1 \, -16 \,x \,-8 \,x^2)^3 }} 
  \Bigr), 
\nonumber 
\end{eqnarray}

\vskip .1cm
 
All these results show a close relation between diagonals of our simple 
examples of rational functions of three variables, and {\em modular forms}. 
This suggests, quite naturally, to perform similar calculations, 
but, now, with {\em four} variables.

\vskip .3cm 

\section{Diagonals of rational functions of four variables }
\label{four}

Let us now consider diagonal of rational functions  
of the form $\, 1/(1 \, -P(x, \, y, \, z, \, w))$ where $\, P(x, \, y, \, z, \, w)$
is a polynomial of four variables $\, x, \, y, \, z, \, w$,  
sum of monomials $\, x^m \, y^n \, z^p \, w^q$
 where the degrees $\, m, \, n, \, p, \, q$
are $\, 0$ or $\, 1$, and where the coefficients in front of these 
monomials are restricted to take {\em only two values} $\, 0, \,  1$.
With these two constraints (on the degrees and on the values of the coefficients),
one finds  an exhaustive list of only $\, 879$ cases\footnote[5]{When different 
rational functions $\, 1/(1 \, -P(x, \, y, \, z, \, w))$ 
yield the {\em same diagonal} (in practice the same first ten 
coefficients of the series) we select one rational function to represent 
the diagonal. This is the way we define these  $\, 879$ different classes
of rational functions.}.

\vskip .1cm 

{\bf Remark 1:} We have used the package ``HolonomicFunctions'' 
written by C. Koutschan~\cite{Koutschan,Koutschan2,RISC}, based on the method of
{\em creative telescoping}~\cite{Zeilberger,Koutschan3}, which enables 
to obtain directly, and very efficiently,
the linear differential operator annihilating the diagonal 
of a given  rational function,
{\em without calculating the series expansion} of the corresponding 
diagonal\footnote[2]{One does not obtain the linear differential operator 
from a ``guessing procedure'' on the series of the diagonal. Furthermore this
algorithm is ``certified'': we are sure that the operator annihilates
the diagonal.}. From time to time Koutschan's algorithms do not provide
the {\em minimal order} linear differential operator. 
It is thus necessary to systematically check whether the linear differential 
operator obtained by creative telescoping is minimal, and if not, 
to find the minimal one.

\vskip .1cm 

{\bf Remark 2:} The command
``FindCreativeTelescoping'', described in~\cite{Koutschan3}, is (usually)
{\em extremely fast}, but it uses some heuristics, which
means that sometimes it can return a non-minimal result or run
forever: it is not an ``algorithm'' in the strict sense\footnote[1]{In contrast, 
Chyzak's algorithm~\cite{Chyzak2} is designed such that
it finds the {\em minimal order operator},
but it is often much slower than Koutschan's heuristics
(in particular for the very large operators emerging 
in physics~\cite{High,Khi6,bo-gu-ha-je-ma-ni-ze-08,bernie2010}). In 
Koutschan's package, Chyzak's algorithm is implemented in the command 
CreativeTelescoping (same input/output specification as FindCreativeTelescoping).}.

\vskip .2cm 

\vskip .1cm 

Among these $\, 879$ cases, the linear differential operators annihilating the 
diagonals have minimal orders running from $\, 1$ to $\, 10$, as given in
Table \ref{tab:table}.

\vskip .1cm

\begin{table}[htdp]
\caption{\label{tab:table}
Number of operators corresponding to the various orders.
 }

\vskip .1cm

\begin{center}
\begin{tabular}{|c|c|c|c|c|c|c|c|c|c|c|c}\hline
\hline
    Order of Oper.& 1 & 2   &  3  &  4 &  5 &  6 &  7  &  8 &  9 &  10   \\   \hline
    Number of Oper. & 1  &  2  &  20 &  128 &  240 &  231 &  155  &  54  &  41  &  7     \\  \hline
 \hline
 \end{tabular}

\end{center}
\end{table}
The order-one linear differential 
operator corresponds to the diagonal of the rational 
function $\, 1/(1 \,-z \,  -w y - x y)$. 
It is easily seen that its diagonal is equal to 1, 
so the order-one operator is just $\, D_x$.

\vskip .1cm

For the order-two linear differential operators,
 the corresponding diagonals 
are two pullbacked $\, _2F_1$ hypergeometric
functions, respectively
\begin{eqnarray}
\label{pu1}
 \hspace{-0.95in}&& \quad \quad    \,   \quad  
(1 \, -32 \,x \, +16\,{x}^{2})^{-1/4} \cdot \, 
 \,_2F_1\Bigl([{{1} \over {12}}, \, {{5} \over {12}}], \, [1], \, P(x)\Bigr) 
\nonumber   \\
\hspace{-0.95in}&& \quad    \quad  \quad  \quad  \quad  \quad 
       \hbox{where:} 
\quad\quad \quad   \quad     
P(x)\, \,\, = \, \, \,  \, \, 
1728\,{\frac {{x}^{3} \cdot \, (2 \,-71\,x \,+16\,{x}^{2}) }{ 
(1 \, -32\,x \, +16\,{x}^{2})^{3}}}, 
\end{eqnarray}
for the diagonal of $\, \, 1/(1-(y\,+z\,+x\, z \,+x\, w \, +x\, y\, w))$
which reads
\begin{eqnarray}
\label{serpu1}
\hspace{-0.95in}&&    \quad   \quad    
1\, +8\,x\, +156\,{x}^{2}\, +3800\,{x}^{3}\, +102340\,{x}^{4}\, +2919168\,{x}^{5}\, 
+86427264\,{x}^{6}
\nonumber \\ 
\hspace{-0.95in}&&  \quad   \quad   \quad   \quad    \quad  \, 
+2626557648\,{x}^{7}\, +81380484900\,{x}^{8} 
\, + 2559296511200\,{x}^{9} \,  \,  \,+\, \, \cdots 
\end{eqnarray}
and
\begin{eqnarray}
\label{pu2}
 \hspace{-0.95in}&& \quad \quad   \quad  \quad    
(1 \, -40\,x \, +16\,{x}^{2})^{-1/4} \cdot \, 
 \,_2F_1\Bigl([{{1} \over {12}}, \, {{5} \over {12}}], \, [1], \, P(x)\Bigr) 
\\
\hspace{-0.95in}&& \quad \quad  \quad  \quad \quad      \quad    \quad 
   \hbox{where:} 
\quad\quad \quad   \quad     
P(x)\, \,\, = \, \, \,  \, \, 
6912\,{\frac {{x}^{3} \cdot \, (1 \, -44\,x \, -16\,{x}^{2}) }{
 (1 \, -40\,x \, +16\,{x}^{2})^{3}}}.
\nonumber   
\end{eqnarray}
for\footnote[1]{These two series (\ref{serpu1}) and (\ref{serpu2})
are not in Sloane's on-line encyclopedia http://oeis.org.} the diagonal of
$\, 1/(1-(y\, +z\,+x\, w \, +x\, z\, w\,  +x\, y\, w))$ which reads:
\begin{eqnarray}
\label{serpu2}
 \hspace{-0.95in}&&    \quad   \quad     
1 \,\, +10\,x \,\, +246\,{x}^{2} \,\, +7540\,{x}^{3} 
\,\, +255430\,{x}^{4} \,\, +9163980\,{x}^{5} \, 
+341237820\,{x}^{6}
\nonumber \\ 
\hspace{-0.95in}&&  \quad    \quad     \quad  \,
 \, +13042646760\,{x}^{7} \, +508236930630\,{x}^{8} \,
 +20101587623260\,{x}^{9}\,  \, + \, \, \, \cdots 
\end{eqnarray}

\vskip .1cm

Note, however, that the  series (\ref{serpu1}) of
the diagonal of $\, 1/(1-(y\,+z\,+x\, z \,+x\, w \, +x\, y\, w))$
actually identifies with the diagonal of a rational function 
of just {\em three} variables $\, 1/(1\, - x - y - z \, -x\,y)$,
already found among the previous $\, 20$ cases of 
section (\ref{three}). 
Similarly the  series (\ref{serpu2}) of
the diagonal of 
$\, 1/(1-(y\, +z\,+x\, w \, +x\, z\, w\,  +x\, y\, w))\,$
also identifies with the diagonal of a rational function 
of three variables $\,\, 1/(1\, - x - y - z \, -x\,y \, - y\, z)$. 

\vskip .1cm
 
The results for all these $\, 879$ cases are 
given exhaustively in our web page of supplementary 
material~\cite{supplementary}. Let us summarize these results
in the following.

\vskip .1cm 

\vskip .1cm 

For all the twenty cases, corresponding to order-three linear differential 
operators, we have $\, SO(3, \, \mathbb{C})$ differential Galois groups. As a 
consequence~\cite{DiffAlgGreen}, {\em all} these linear differential
operators are actually 
{\em symmetric squares of order-two operators}, some with very 
simple $\, _2F_1$ hypergeometric 
functions, namely $\, _2F_1([3/8, 1/8], \,[1], \,256\,x^3)$ or 
$\, _2F_1([1/3, 1/6], \,[1], \, 108\,x^3)$, some with, at first sight, 
more involved HeunG function solutions~\cite{IsingModularForms}
 which turn out to be pullbacked
 $\, _2F_1$ hypergeometric functions, with {\em two possible pullbacks}, 
and, in fact, {\em modular forms}~\cite{IsingModularForms}. 

\vskip .1cm  

The  $\, 128$  order-four linear differential operators are (non-trivially)
 homomorphic to their adjoints. They have  $\, SO(4, \, \mathbb{C})$ differential 
Galois groups and have a canonical decomposition~\cite{Canonical} 
of the form $\, (A_1 \, B_3 \, + \, 1) \cdot \, r(x)$, where 
$\, A_1$ and $\, B_3$ are respectively order-one and {\em order-three
self adjoint} operators, $\, r(x)$ being a rational function.

\vskip .1cm 

Similarly, the $\, 240$ order-five linear differential operators are (non-trivially)
 homomorphic to their adjoints. They have  $\, SO(5, \, \mathbb{C})$ differential 
Galois groups and have a canonical decomposition~\cite{Canonical} 
of the form $\, (A_1 \, B_1 \, C_3 \, + A_1 \,+ \, C_3) \cdot \, r(x)$, where 
$\, A_1$, $\, B_1$ and $\, C_3$ are respectively two order-one and one 
{\em order-three self adjoint} operators, $\, r(x)$ being a rational function.

\vskip .1cm 
 
The $\, 231$ order-six linear differential operators are (non-trivially)
 homomorphic to their adjoints. They have  $\, SO(6, \, \mathbb{C})$ differential 
Galois groups and have a canonical decomposition~\cite{Canonical} 
of the form 
$\, (A_1 \, B_1 \, C_1 \, D_3 \, + A_1 \, B_1\, $
$+ \,A_1 \, D_3 \,+ \, C_1 \, D_3 \,+ \,1) \cdot \, r(x)$, where 
the $\, A_1, \,B_1, \,C_1, \, $ operators are order-one self-adjoint operators,
the rightmost operator $\, D_3$ being an {\em order-three self-adjoint} 
operator.

\vskip .1cm 

 The following $\, 155$ order-seven linear differential operators are (non-trivially)
 homomorphic to their adjoints. They have  $\, SO(7, \, \mathbb{C})$ differential 
Galois groups and have a canonical decomposition~\cite{Canonical} 
of the form 
$\, (A_1 \,B_1 \,C_1 \,D_1 \,E_3 +A_1\,B_1 \,E_3\,
 +A_1 \,D_1 \,E_3 \,+A_1\,B_1 \,C_1\,$
$ +C_1  \,D_1  \,E_3 \, +A_1\,+C_1\, +E_3) \cdot \, r(x)$, where 
the $\, A_1, \,B_1, \,C_1, \,D_1 $ operators are order-one self-adjoint operators,
the rightmost operator $\, E_3$ being an order-three self-adjoint 
operator.

\vskip .1cm 

The $\, 54$ order-eight linear differential operators are (non-trivially)
 homomorphic to their adjoints. They have  $\, SO(8, \, \mathbb{C})$ differential 
Galois groups and have a canonical decomposition
described in~\cite{Canonical}, generalization
of the previous ones,  with, again, five order-one self-adjoint operators,
and a  rightmost {\em order-three self-adjoint} operator.

\vskip .1cm 

The $\, 41$ order-nine linear differential operators are (non-trivially)
 homomorphic to their adjoints. They have  $\, SO(9, \, \mathbb{C})$ differential 
Galois groups and have a canonical decomposition
described in~\cite{Canonical}, generalization
of the previous ones,  with, six order-one self-adjoint operators,
and a  rightmost {\em  order-three self-adjoint} operator.

\vskip .1cm 

Finally, the seven order-ten linear differential
operators are (non-trivially)
 homomorphic to their adjoints. They have  $\, SO(10, \, \mathbb{C})$ differential 
Galois groups and have a canonical decomposition
described in~\cite{Canonical}, generalization
of the previous ones,  with, seven order-one self-adjoint operators,
and a  rightmost {\em order-three self-adjoint} operator.

\vskip .1cm 

These results are reminiscent of the results obtained on 
a set of 210  explicit linear differential operators annihilating
 diagonals of rational functions, arising from  mirror symmetries
 and corresponding to reflexive 4-polytopes~\cite{Lairez},
recalled in the introduction. One notes, however, that 
the {\em symplectic} $\, Sp(n, \, \mathbb{C})$ differential Galois groups
with a canonical decomposition in order-two
 self-adjoint operators and a rightmost  order-four self-adjoint 
operator, encountered with these reflexive 4-polytopes
examples, is now replaced by {\em  orthogonal}  $\, SO(n, \, \mathbb{C})$ 
differential Galois groups with a canonical decomposition in order-one
 self-adjoint operators and a rightmost  {\em order-three self-adjoint} 
operator.

\vskip .1cm 

The calculations performed here, in order to see that these 
(quite large) linear differential operators are (non trivially) homomorphic
to their adjoints and to find their canonical decompositions~\cite{Canonical},
are similar to the ones described in~\cite{Canonical} for the  
reflexive 4-polytopes examples: in order to find the intertwiner, we 
introduced a specialized algorithm because the Maple 
command $\, Homomorphisms(adjoint(L),\,L)$ never terminates  
on these large operators~\cite{Canonical}.
We use a fuchsian linear differential system associated to $\, L$, 
the {\em theta-system}~\cite{Canonical}, a slight 
generalization of the companion system
which has simple poles at each finite singularity. 
One then finds a rational solution of an associated system 
with similar coefficients (its second symmetric/exterior power), 
which gives the intertwiner.
An inversion of this intertwiner modulo $\, L$ gives the intertwiner 
corresponding to the Maple command $\, Homomorphisms(L, \, adjoint(L))$. 
Finally, one obtains  the canonical decomposition from simple euclidean 
divisions~\cite{Canonical}. 

\vskip .1cm 

For the last seven order-ten operators even the theta-system calculations 
were quite massive, 
and required up to two weeks of CPU time and 
 up to $\, 80$ Gigaoctets of memory for one linear differential operator. 

\vskip .1cm 

\vskip .2cm 

\section{Diagonals of rational functions associated with orthogonal 
as well as symplectic groups}
\label{orthosymplect}

It is tempting to simply, and straightforwardly, 
generalize\footnote[1]{In particular after 
a set of other unpublished results we have obtained.} these two sets 
of results for  diagonals of  well-defined finite sets of rational 
functions, namely the symplectic $\, Sp(n, \, \mathbb{C})$
differential Galois groups~\cite{Lairez} (with an order-four 
rightmost self-adjoint operator as for  
reflexive 4-polytopes~\cite{Lairez}),
and the orthogonal  $\, SO(n, \, \mathbb{C})$ 
differential Galois groups (with  a rightmost  order-three 
self-adjoint operator). The situation can be
slightly more involved (and richer) than a straightforward 
generalization of the previous results. In fact, {\em diagonals are 
not necessarily} solutions of an {\em irreducible} linear 
differential operator.

To see this let us consider the diagonal\footnote[5]{Note that this diagonal 
is factorized but is not the Hadamard product of two diagonals 
(see \ref{morex}).} of the rational function of four variables 
with a factorized denominator
$\, R \, = \, \,$
$ 1/(1 - x \, -y \, -u \, -z)/(1 - u  - z  -u\, z)$, which  
reads:
\begin{eqnarray}
\label{diagR}
\hspace{-0.95in}&&  \, \, \,  \, 
Diag(R) \, = \, \, \, \, 1 \,  \,\,  +42\, x  \,\,  +4878 \, x^2  \, 
+748020 \, x^3 \,  +130916310 \, x^4 \,  +24762428460 \, x^5 
\nonumber \\ 
\hspace{-0.95in}&&  \quad \quad \quad \quad
 \, +4929691760532 \, x^6 \,\, 
  +1017691904736936 \, x^7 \,  \,
+  \, \, \cdots 
\end{eqnarray}
It is solution of an order-seven linear differential operator which 
factorizes\footnote[3]{Using the DFactorLCLM command one sees that one 
does not have a direct sum factorization, just the simple factorization
$\, L_7 \, = \, \, L_3 \cdot \, L_4$.} 
into an {\em order-three} and an {\em order-four} operator: 
$\, L_7 \, = \, \, L_3 \cdot \, L_4$. Note that $\, L_4$ {\em does not annihilate}
 this diagonal. 

\vskip .1cm 

This means that the series (\ref{diagR}) is solution of 
$\, L_7 \, = \, \, L_3 \cdot \, L_4$, but not 
of $\, L_4$ which has a solution analytic at $\, x\, = 0$, with a series 
expansion with {\em integer coefficients}\footnote[8]{A natural question corresponds
 to ask if such series (\ref{L4ser}) with integer coefficients are necessarily a diagonal
of rational function. It is true for ODEs of minimum weight for the monodromy 
filtration~\cite{Christol}, but in the general case, it
 is still a conjecture. It, however, seems that
one can prove that such series are ``automatic'' (i.e. reduce 
to algebraic functions) modulo powers of primes (G. Christol, 
 private communication).} 
different from (\ref{diagR}):
\begin{eqnarray}
\label{L4ser}
\hspace{-0.95in}&&  \quad \quad 
1\,\, +214\,x\,\, +97278\,{x}^{2}\,+53983020\,{x}^{3}\,
+32898451110\,{x}^{4}\,\,  +21172639875156\,{x}^{5}
\nonumber \\
\hspace{-0.95in}&&  \quad \quad \quad  \quad \quad \quad 
\,+14121624413802444 \,{x}^{6}\, \,+ \,\, \cdots 
\end{eqnarray}
The series  $\,  \, L_4(Diag(R))$, solution of $\, L_3$, is 
{\em globally bounded}~\cite{Short,Big}.
If one normalizes $\, L_4$ to be an operator with polynomial coefficients 
(instead of being a monic operator), the  
series  $\,  \, L_4(Diag(R))$, solution of $\, L_3$, is a series
with {\em integer coefficients}: 
\begin{eqnarray}
\label{PolL4ser}
\hspace{-0.95in}&&   
17888\, +25769200\,x\,  +17312032256\,{x}^{2}\, +8722773606816\,{x}^{3} 
\, +3775743401539200\,{x}^{4}\, 
\nonumber \\
\hspace{-0.95in}&&  \quad \quad \quad  \quad 
+1486619414765913792\,{x}^{5}\,  + 548416028673746513280\,{x}^{6}
\,\, + \,\, \cdots 
\nonumber 
\end{eqnarray}

Note that the order-three operator $\, L_3$ has 
a canonical $\, SO(3, \, \mathbb{C})$ orthogonal decomposition~\cite{Canonical}
$\, (A_1 \, B_1 \, C_1 \, + A_1 \,+ \, C_1) \cdot \, r(x)$,
where $\, A_1$, $\, B_1$,  $\, C_1$,  are order-one self-adjoint operators.
In contrast  $\, L_4$ has the canonical $\, Sp(4, \, \mathbb{C})$ symplectic 
decomposition $\, (A_2 \, B_2 \, +1) \cdot \, \rho(x)$,
where $\, A_2$, $\, B_2$ are order-two self-adjoint operators.
In other words, the diagonal (\ref{diagR}) is associated with both 
orthogonal {\em and} symplectic differential Galois groups! This is, in fact, 
the situation we expect generically: the diagonal of the rational function
will be solution of a {\em non-irreducible} linear differential operator,
{\em each of its factors} corresponding to {\em orthogonal or symplectic}
 differential Galois groups. 

\vskip .1cm 

We have, for instance, the same results for another rational function
such that its denominator is factorised.
The diagonal of  
$\, R \, = \, \, 1/(1 - x - y -z -u)/(1 -u - z  - x\, z )$,
namely 
\begin{eqnarray}
\label{DiagL4ser2}
\hspace{-0.95in}&&   \quad  \quad 
1 \, \,\, +44\,x \,\, \, +5061\,{x}^{2} \,\,\,
  +771000\,{x}^{3} \,\,  +134309890\,{x}^{4} \,
 +25316919264\,{x}^{5} 
\nonumber \\
\hspace{-0.95in}&&   \quad  \quad  \quad  \quad   \quad 
\, +5026804760628\,{x}^{6} \,\,  \, +  \,  \, \cdots 
\end{eqnarray}
is solution of an order-seven linear differential operator which 
factorizes into an {\em order-three} and 
an {\em order-four} operator: 
$\, L_7 \, = \, \, L_3 \cdot \, L_4$ (but, again there is no 
direct sum factorization). 
This diagonal (\ref{DiagL4ser2}) is solution of $\, L_7$
but not of  $\, L_4$.  If one normalizes $\, L_4$ to be 
an operator with polynomial coefficients 
(instead of being a unitary operator), the  
series  $\,  \, L_4(Diag(R))$, solution of $\, L_3$, is a series
with {\em integer coefficients}: 
\begin{eqnarray}
\label{PolL4ser3}
\hspace{-0.95in}&&    \quad  \quad 
16\,\, -94464\,x\,\, -127052100\,{x}^{2}\,\, -86146838400\,{x}^{3}\,\,
-44244836836200\,{x}^{4}
\nonumber \\
\hspace{-0.95in}&&  \quad \quad \quad  \quad   \, \, 
-19495756524980736\,{x}^{5}\, -7791904441995369696\,{x}^{6}
\,\,\, + \,\, \cdots 
\end{eqnarray}
The order-three linear differential operator $\, L_3$  is 
MUM (maximal unipotent monodromy~\cite{Todorov}) 
and has a canonical $\, SO(3, \, \mathbb{C})$ decomposition
$\, (A_1 \, B_1 \, C_1 \, + A_1 \,+ \, C_1) \cdot \, r(x)$,
where $\, A_1$, $\, B_1$,  $\, C_1$,  are order-one self-adjoint operators.
It is not the symmetric square of an order-two operator, it is homomorphic to the 
 symmetric square of an order-two operator. 
In contrast  $\, L_4$ has the canonical $\, Sp(4, \, \mathbb{C})$ decomposition
$\, (A_2 \, B_2 \, +1) \cdot \, \rho(x)$
where $\, A_2$, $\, B_2$ are order-two self-adjoint operators.

\vskip .1cm 

\section{Diagonals of rational functions 
corresponding to $\, _nF_{n-1}$ hypergeometric functions}
\label{nFn}

Beyond these finite sets of examples of diagonals of rational functions, 
 namely the 210 reflexive 4-polytopes operators~\cite{Lairez}
with symplectic $\, Sp(n, \, \mathbb{C})$
differential Galois groups,
and these $\, 879$  operators of section (\ref{four})
associated with diagonals of rational 
functions of four variables with
 orthogonal  $\, SO(n, \, \mathbb{C})$ 
differential Galois groups, one can find infinite families of 
diagonals of rational functions for which exact results 
can be obtained corresponding to $\, _nF_{n-1}$ hypergeometric functions.

For instance, the diagonal of the rational function of three variables 
$\, R \, = \, \, 1/(1 - x - z - y^n)$,  with $\, n$ being a positive integer,
is a  $\, _{2\, n}F_{2\, n  \, -1}$ hypergeometric function
\begin{eqnarray}
\hspace{-0.95in}&&  \quad \quad 
Diag\Bigl( {{1} \over { 1 - x - z - y^n}}\Bigr)  
\,   \, \, = \, \,  \,  
\nonumber \\
\hspace{-0.95in}&& \quad \quad \quad  \quad 
_{2\, n}F_{2\, n  \, -1}\Bigl([{{1} \over {2\, n \, +1}}, \, {{2} \over {2\, n \, +1}}, \,
{{3} \over {2\, n \, +1}}, \, \cdots, \, {{2  \, n} \over {2\, n \, +1}}],
 \\
\hspace{-0.95in}&& \quad \quad \quad  \quad  \quad \quad    \,
[{{1} \over {n}}, \, {{1} \over {n}}, \,
{{2} \over {n}}, \, {{2} \over {n}}, \, 
{{3} \over {n}}, \, {{3} \over {n}}, \, \cdots, \,{{n\, -1} \over {n}},
\, {{n\, -1} \over {n}}, \, {{n} \over {n}}], \, \, \, 
 {{ (2\, n \, +1)^{2\, n \, +1} } \over {n^{2\, n}}} \cdot \, x^n   \Bigr).  
\nonumber 
\end{eqnarray}
The corresponding linear differential operator is an order-$2n$ 
linear differential operator having a {\em symplectic} 
$\, Sp(2\, n, \, \, \mathbb{C})$ differential Galois group.

The diagonal of the rational functions of three variables 
$\, R \, = \, \, 1/(1 - x - z \, - x^n \,y^n)$,  
with $\, n$ being a positive integer,
is a  $\, _{n}F_{n  \, -1}$ hypergeometric function
\begin{eqnarray}
\hspace{-0.95in}&&  \quad \quad \quad \quad 
Diag\Bigl( {{1} \over { 1 - x - z \, - x^n \, y^n}}\Bigr)  
\,   \, = \, \,  \,  
\nonumber \\
\hspace{-0.95in}&& \quad \quad \quad  \quad \quad \quad \quad 
_{n}F_{n  \, -1}\Bigl([{{1} \over {n \, +1}}, \, {{2} \over {n \, +1}}, \,
{{3} \over {n \, +1}}, \, \cdots, \, {{n} \over {n \, +1}}],
 \\
\hspace{-0.95in}&& \quad \quad \quad  \quad \quad  \quad \quad  \quad \quad  \,
[{{1} \over {n}}, \, {{2} \over {n}}, \,
{{3} \over {n}},  \, \cdots, \,{{n\, -1} \over {n}}], \, \, \, 
{{ (n \, +1)^{n \, +1} } \over { n^{\, n}}} \cdot \, x^n   \Bigr).   
\nonumber 
\end{eqnarray}
The corresponding linear differential operator is an order-$n$ 
linear differential  operator having an 
{\em orthogonal} $\, O(n, \, \, \mathbb{C})$ differential Galois group.

The diagonal of the rational functions of three variables 
$\, R \, = \, \, 1/(1 - x - z \, - x \,y^n)$,  with $\, n$ being a positive integer,  
is a  $\, _{n}F_{n  \, -1}$ hypergeometric function.
For $\,n$  even it reads:
\begin{eqnarray}
\label{xyn}
\hspace{-0.95in}&&  \quad \quad \quad 
Diag\Bigl( {{1} \over { 1 - x - z \, - x \, y^n}}\Bigr)  
\,   \,\,  = \, \,  \,  
\nonumber \\
\hspace{-0.95in}&& \quad \quad \quad \quad  \quad \quad 
_{n}F_{n  \, -1}\Bigl([{{1} \over {2\, n}}, \, {{3} \over {2\, n}}, \,
{{5} \over {2\, n}}, \, \cdots, \, {{2\,n \, -1} \over {2\, n}}],
 \\
\hspace{-0.95in}&& \quad \quad \quad  \quad \quad \quad  \quad  \quad  \,
[{{1} \over {n \, -1}}, \, {{2} \over {n\, -1}}, \,
{{3} \over {n\, -1}},  \, \cdots,
 \,{{n\, -2} \over {n\, -1}}, \,{{n\, -1} \over {n\, -1}}], 
\, \, \,  {{ 4^{n} \cdot \, n^n} \over {(n \, -1)^{n \, -1}}}
 \cdot \, x^n   \Bigr).   
\nonumber 
\end{eqnarray}
The corresponding linear differential operator is an order-$n$ 
linear differential operator having a 
{\em symplectic} $\, Sp(n, \, \, \mathbb{C})$ differential Galois group.

For $\,n$  odd the same formula (\ref{xyn}) holds. Note, however, that
the argument $\, 1/2$ appears in both the first and second list of arguments of the
$\, _{n}F_{n  \, -1}$ hypergeometric function, and, hence, can be avoided,
which is thus a $\, _{n-1}F_{n  \, -2}$ hypergeometric function.
The corresponding linear differential operator is an order-$(n-1)$ 
operator having a {\em symplectic} $\, Sp(n-1, \, \, C)$ differential 
Galois group.

\vskip .1cm 

\section{Diagonals of rational functions associated with operators 
non homomorphic to their adjoints}
\label{nonhomo}

After all this accumulation of examples of diagonals of rational functions
associated with orthogonal, or symplectic, differential Galois groups, 
it is tempting to conjecture that counter-examples\footnote[3]{Given in 
the introduction corresponding to a $\, SL(3, \, \mathbb{C})$  
differential Galois group, and thus the corresponding operator 
cannot be homomorphic to its adjoint {\em even with an algebraic extension}.} 
like $\,_3F_2([1/3, \, 1/3, \, 1/3], \, [1, \, 1], \,  3^6 \, x)$,
are quite ``rare exceptional cases'' that one 
can easily detect, and, hopefully, understand, as situations 
of  algebraic varieties ``sufficiently singular'' 
{\em to break the Poincar\'e duality}~\cite{Griffiths}. 
Along this line of seeking for diagonals of rational functions with 
a ``sufficiently singular'' denominator, let us try to provide examples of 
operators that are {\em not homomorphic to their adjoints}. 

\vskip .1cm 

\ref{attempt} provides two attempts to find examples of 
diagonals of rational functions such that their corresponding linear
differential operators would not be homomorphic to their 
adjoints. We first study in \ref{Arnold} a set of singular denominators 
for the rational functions, the polynomials of three variables
 corresponding to {\em classifications of singular varieties} 
performed by V. I. Arnold~\cite{Arnold}. {\em All the corresponding 
operators yield symplectic $\, Sp(n, \, \mathbb{C})$ differential Galois groups}: 
 the kind of singular behaviour required to 
``break the Poincar\'e duality'',
considered in V. I. Arnold~\cite{Arnold} for three variables, 
is not ``sufficiently singular''. 

\vskip .1cm 

It could be that these ``sufficiently singular'' situations require 
much {\em more than three variables}\footnote[1]{Counter-examples, like 
$\,_3F_2([1/3, \, 1/3, \, 1/3], \, [1, \, 1], \,  3^6 \, x)$,
correspond to diagonal of algebraic functions of three variables (which are
very simple since they are products of the algebraic functions of one variable),
but this means that they are  diagonals of rational functions of, at first sight,
 {\em six} variables.}. Therefore we have considered, here, rational functions 
of {\em six} variables, and since the theory of singularities of algebraic 
varieties~\cite{Europe,Hironaka}
suggests that the situation of algebraic varieties ``singular enough'' often 
correspond to denominators that {\em factor}\footnote[9]{This is actually the case
for the counter-examples, like 
$\,_3F_2([1/3, \, 1/3, \, 1/3], \, [1, \, 1], \,  3^6 \, x)$ 
mentioned in the introduction.}, we have considered,
in \ref{morex}, rational functions  
of the form $\, 1/\left((1\, -P_1)\,(1\, -P_2)\right)$ 
where $\, P_1$ and $\, P_2$ are 
two polynomials of two different sets of three variables\footnote[2]{This 
also corresponds to considering the {\em Hadamard product
of the  diagonal of  $\, 1/(1 \, -P_1(x, \, y, \, z))$ 
and of the  diagonal of  $\, 1/(1 \, -P_2(u, \, v, \, w))$}.}. 
Again we found that {\em all}\footnote[8]{Even using Koutschan's creative 
telescoping program~\cite{Koutschan}, two operators among the $\, 170$ 
were difficult to obtain. Among these $\, 170$
operators, $\, 71$ are so large that they cannot be analyzed
using the DEtools commands, and required to switch to a theta-system
approach (see~\cite{Canonical}). 
} the operators that could be
analyzed among the $\, 170$ operators
in \ref{morex}, are {\em homomorphic to their adjoints,
and have  symplectic or orthogonal differential Galois groups}. 

\vskip .1cm 

We also considered the (minimal order) 
linear differential operators annihilating large sets\footnote[5]{Some 
miscellaneous heuristic examples are given in the supplementary 
materials~\cite{supplementary}.} of 
{\em Hadamard products}~\cite{Short,Big,Hadamard}
 {\em of algebraic functions}, and, {\em each time},  obtained 
that these linear differential operators are homomorphic to their adjoints.
Surprisingly, finding diagonals of rational functions
such that their annihilating linear differential operators are {\em not} 
homomorphic to their adjoints, is not so easy, examples like 
$\,_3F_2([1/3, \, 1/2, \, 1/2], \, [1, \, 1], \,  12^2 \, x)$,
being {\em quite rare}. Let us revisit these simple 
hypergeometric examples yielding 
operators that are {\em not homomorphic} to their adjoints. 
  
\vskip .2cm 

\subsection{Examples of diagonals of rational functions with 
no homomorphism to the adjoint}
\label{nonhomosub}

The hypergeometric function 
$\,_3F_2([1/3, \, 1/2, \, 1/2], \, [1, \, 1], \,  12^2 \, x)$,
actually corresponds to a $\, SL(3, \, \mathbb{C})$  differential Galois group. It can be 
seen as the diagonal of a rational function of {\em six variables}
\begin{eqnarray}
\label{diag3F2a}
\hspace{-0.95in}&&  \quad \, \, 
Diag\Bigl(  
{{ 1 \, -9 \, x \, y} \over { (1 - 3\, y - 2\, x + 3\, y^2 + 9\, x^2\, y) 
\cdot \, (1\, - u \,- z) \cdot \, (1\, - v \,- w)}} 
\Bigr)
\\
\hspace{-0.95in}&&  \quad \, \quad  \quad \, = \, \, \, \,\, 
1\,\, \, +12\,x\,\,  +648\,{x}^{2}\,\,  +50400\,{x}^{3}\,\,
  +4630500\,{x}^{4}\, \, +468087984\,{x}^{5}\, 
\,\, + \, \,\, \cdots 
\nonumber 
\end{eqnarray}
The corresponding (order-three) linear differential operator is 
{\em not homomorphic to its adjoint}, even with an algebraic extension. 
This diagonal  (\ref{diag3F2a}) is of the form 
$\, Diag(R_1(x, \, y) \cdot \, R_2(u, \, z) \cdot \,  R_3(v, \, w))$,
(where $\, R_1$, $\, R_2$ and $\, R_3$ are simple rational functions).
It is, thus, the Hadamard product~\cite{Short,Big,Hadamard} of the three 
diagonals $\,  Diag(R_1(x, \, y))$,  $\,  Diag(R_2(u, \, z))$ and  
$\,  Diag(R_2(v, \, w))$, which are simple {\em algebraic functions},
 respectively:
\begin{eqnarray}
\label{diag3F2a1}
\hspace{-0.95in}&&  \quad \quad  \quad  \, 
Diag\Bigl(  
{{ 1 \, -9 \, x \, y} \over { 1  - 2\, x - 3\, y \, + 3\, y^2 + 9\, x^2\, y }} 
\Bigr) \, \, = \, \, \,\,    {{1} \over { (1 \, -9 \, x)^{1/3}}},  
\end{eqnarray}
\begin{eqnarray}
\label{diag3F2a23}
\hspace{-0.95in}&&  \quad  \quad \quad  \, 
Diag\Bigl(  
{{ 1} \over { 1\, - u \,- z }} 
\Bigr) \, \, = \, \, \, \,  Diag\Bigl(  
{{ 1} \over { 1\, - v \,- w }} 
\Bigr) \, \, = \, \, \, \,  {{1} \over { (1 \, -4 \, x)^{1/2}}}.
\end{eqnarray}

The linear differential operators, annihilating the diagonals (\ref{diag3F2a1})
and  (\ref{diag3F2a23}), are very simple {\em order-one} linear differential operators.
There is, of course, no relation between their differential Galois groups
and the  $\, SL(3, \, \mathbb{C})$ differential Galois group of the order-three 
operator annihilating their Hadamard product (\ref{diag3F2a}): {\em the Hadamard product
does not preserve algebraic structures}, like the differential Galois group.

\vskip .2cm 

{\bf Remark:} Let us consider the following rational function of six variables: 
\begin{eqnarray}
\label{NoverD}
\hspace{-0.95in}&&  \quad \quad 
{{N(x, \, y, \, z, \, u, \, v, \, w)} \over {
(1  -2\,x-3\,y   \, +9\,{x}^{2}y \, +3\,{y}^{2}) 
\cdot \, (1\,-z\,-u)\cdot \, (1 \, -v\,-w)  }} 
\\
\hspace{-0.95in}&&  \quad \quad \quad \quad \quad \quad  
\, \, = \, \, \,  \,  \, 
 {\frac {1 \, -9\,xy}{1  -2\,x-3\,y   \, +9\,{x}^{2}y \, +3\,{y}^{2}}}
\, \,\, +\,\,  \, {{1} \over {1\,-z\,-u}} 
\,\, \, + \, \,  \, {{1} \over {1 \, -v\,-w}}. 
\nonumber 
\end{eqnarray}
It is straightforward to see that the diagonal of this rational 
function (\ref{NoverD}) of six variables is nothing but the sum of 
the three diagonals (\ref{diag3F2a1}) and  (\ref{diag3F2a23}), namely: 
\begin{eqnarray}
\label{sumdiag3F2a23}
\hspace{-0.95in}&&  \quad  \quad \quad \quad \quad \quad \quad  \, 
 {{1} \over { (1 \, -9 \, x)^{1/3}}} \, \, + \, \, {{2} \over { (1 \, -4 \, x)^{1/2}}}.
\end{eqnarray}
This diagonal is solution of an order-two linear differential operator which
is (obviously) a direct sum:
 \begin{eqnarray}
\label{sumdiag3F2a}
\hspace{-0.95in}&&  \quad  \quad \quad  \quad \quad \quad  \, 
 \Bigl(D_x \, - \, \,   {{3} \over { 1 \, -9 \, x}} \Bigr)
  \, \oplus  \, \, \Bigl(D_x \, - \, \, {{2} \over { 1 \, -4 \, x}} \Bigr).
\end{eqnarray}
The differential Galois group of the (order-three) operator that annihilates 
(\ref{diag3F2a}), and the one of the (order-two) operator (\ref{sumdiag3F2a}) 
that annihilates (\ref{NoverD}), are quite different, 
{\em even if the denominators of} (\ref{diag3F2a}) and (\ref{NoverD}) 
{\em are the same}. In a theory of singularity 
perspective~\cite{Hironaka,Europe}, the statement that the 
differential Galois group depends essentially 
on the denominator\footnote[2]{Mathematicians would say that
changing the numerator may change the $n$-form one integrates, that 
one must consider the Gauss-Manin Picard-Vessiot module of these 
differentials. We just try here, heuristically, to make crystal clear that 
numerators matter.
} of the rational function that ``encodes the singularities'',
has to be taken ``cum grano salis''. The restriction we have imposed on our rational 
functions by imposing the numerators to be equal to one, is far from innocent.  

\vskip .2cm 

\subsection{More examples of diagonals of rational functions 
with no homomorphism to the adjoint }
\label{morenonhomo}

A slight modification of the previous example (\ref{diag3F2a}) amounts 
to considering the diagonal of a rational function of {\em six variables}
\begin{eqnarray}
\label{diag3F2b}
\hspace{-0.95in}&&  \quad  \quad  \quad \, 
Diag\Bigl(  
{{ 1 \, -9 \, x \, y} \over { (1 - 3\, y - 2\, x + 3\, y^2 + 8\, x^2\, y) 
\cdot \, (1\, - u \,- z) \cdot \, (1\, - v \,- w)}} 
\Bigr),  
\end{eqnarray}
where the coefficient of $\, x^2 \, y$ has been modified ($9\, x^2\, y$
changed into $\, 8\, x^2\, y$).
The corresponding (order-five) linear differential operator is 
{\em not homomorphic to its adjoint}, even with an algebraic extension,
and its differential Galois group is  $\, SL(5, \, \mathbb{C})$.

\vskip .1cm 

Another slight modification of (\ref{diag3F2a})
 amounts to considering  the diagonal of a rational function 
of {\em five variables}
\begin{eqnarray}
\label{diag3F2c}
\hspace{-0.95in}&&  \quad  \quad  \quad \, 
Diag\Bigl(  
{{ 1 \, -9 \, x \, y} \over { (1 - 3\, y - 2\, x + 3\, y^2 + 9\, x^2\, y) 
\cdot \, (1\, - u \, -v\,  \,- z \,- w)}} 
\Bigr).  
\end{eqnarray}
The corresponding (order-four) linear differential operator is 
{\em not homomorphic to its adjoint}, even with an algebraic extension,
and its differential Galois group is $\, SL(4, \, \mathbb{C})$.

\vskip .1cm 

More examples of diagonals of rational functions with linear differential 
operators, that are not homomorphic to their adjoints, are given in \ref{misnon}.

\vskip .2cm 

{\bf To sum-up:} The cases such that the diagonals of 
rational functions do not yield linear differential operators homomorphic 
to their adjoints, are far from being understood, either from a differential algebra
viewpoint, or from a {\em theory of singularity} perspective. From an experimental
mathematics perspective, what we see is that  diagonals of 
rational functions ``almost systematically'' (but not always !)
 yield linear differential operators 
homomorphic to their adjoints, thus giving selected differential Galois groups.
Diagonals of rational functions were seen~\cite{Short,Big} to naturally emerge
in physics. However, in a mathematical framework,  {\em not related to physics},
diagonals of rational functions seem to ``almost systematically''
yield {\em orthogonal or symplectic} groups, with, at first sight, no obvious
``physical interpretation''.

\vskip .2cm 

\section{Conclusion}
\label{conclusion}

We have introduced well-defined sets of
diagonals of rational functions of three, four and six variables, showing 
that all these examples yield 
selected differential Galois groups, namely {\em orthogonal and symplectic} groups.

It has been seen that, in our set of rational functions of {\em three}
variables with denominators with degree bounded by $1$, 
all diagonals correspond to {\em modular forms} 
that can all be written
as $\, _2F_1$ hypergeometric functions with 
{\em two pull-backs}~\cite{IsingModularForms}, related by a 
{\em modular curve}.

\vskip .1cm 

We have seen that a set of $\, 879$ diagonal of rational functions
of {\em four} variables correspond to {\em orthogonal} $\, SO(n, \, \mathbb{C})$
differential Galois groups with a remarkable canonical 
decomposition~\cite{Canonical} with a rightmost self-adjoint operator of order three. 
These results were obtained using the very powerful package ``HolonomicFunctions'' 
written by C. Koutschan~\cite{Koutschan,Koutschan2,RISC}, based on the method of
{\em creative telescoping}~\cite{Zeilberger}, which enables to obtain directly, 
and very efficiently, the linear differential operator 
annihilating a given diagonal of a rational function,
without calculating the series expansion of the corresponding diagonal. 
In order to find the homomorphisms of these operators to their adjoints,
which is the first step towards the analysis of the differential Galois groups 
of these operators and their ``canonical decompositions''~\cite{Canonical},
we have, for large linear differential operators, also used a new algorithm 
that requires to work on the linear {\em theta-system}
associated with the operators~\cite{Canonical}. 

We have also seen (\ref{morex} below) that $\, 170$ diagonals 
of  rational functions of 
six variables of the form 
$\, 1/\left( (1 \, -P_1(x, \, y, \, z)) (1 \, -P_2(u, \, v, \, w)) \right)$,
actually correspond to a quite rich set of linear 
differential operators with orthogonal or symplectic  differential 
Galois groups. The systematic analysis performed in this paper
of these three sets of diagonals of rational functions of respectively three, 
four and six variables, suggests that diagonals of rational functions
``almost systematically'' yield orthogonal or symplectic  differential 
Galois groups.

\vskip .1cm 

A contrario, we have provided in section (\ref{nonhomo}) a few  
miscellaneous examples of diagonals of rational functions 
where the corresponding linear differential operators are 
{\em not homomorphic to their adjoints}. These cases, such 
that the diagonals of rational functions do not yield linear 
differential operators homomorphic 
to their adjoints, are far from being fully understood. Is
it possible that such cases could also emerge with the 
diagonals of rational functions appearing in physics? This 
remains an open and challenging question. 

\vskip .1cm 
\vskip .1cm 

\vskip .1cm

{\bf Acknowledgments.} 
J-M. M. would like to thank R. J. Baxter for so many open, friendly
scientific discussions during these last 35 years, 
since he met him in the 1980 Enschede Summer School.  
J-M. M. would like to thank L. Dumont and P. Lairez for many enlightening 
discussions on diagonals of rational functions and C. Koutschan for
fruitful comments on his ``HolonomicFunctions'' package.
S. B. would like to thank the LPTMC and the CNRS for kind support. 
We would like to thank Y. Abdelaziz for checking 
many of our results. We thank the Research Institute for Symbolic Computation,
(Peter Paule) for access to the RISC HolonomicFunctions package. We thank
G. Christol for many enlightening discussions on Poincar\'e duality.
As far as physicists authors are concerned, this work has been performed 
without any support of the ANR, the ERC or the MAE, or any PES of the CNRS. 

\vskip .1cm 

\vskip .3cm 
\vskip .2cm 

\appendix

\section{Diagonals of rational functions of three variables:
 some modular forms }
\label{modular}

Let us recall the two {\em Hauptmoduls}~\cite{IsingModularForms}
$\, 12^3/j_3$ and $\, 12^3/j_4$ of Table 4 of Maier~\cite{Maier1}:
\begin{eqnarray}
\label{Maierrecall}
\hspace{-0.95in}&& \quad  \quad   
H_3(z) \,\, = \, \,\,
{{ 12^3 \cdot \, z} \over { (z \, +27)\cdot \, (z \, +3)^3}}, 
 \quad  \quad 
 H_4(z) \,\, = \, \,\,
{{ 12^3 \cdot \, z \cdot (z+16)} \over { (z^2\, +16\, z \, +16)^3}}.
\nonumber 
\end{eqnarray}

\vskip .1cm 

$\bullet$ The diagonal of $\, 1 / (1\, - x - y - y\, z \, -x\,z)$
{\em as well as the  diagonal of} $\, 1 / (1\, - x - y - \, z \, +x\,z)$,
correspond to the sequence 
$[1, 4, 36, 400, 4900, 63504, \, \cdots]$ of the 
complete elliptic integral $\, K(4\, x^{1/2})$
(oeis number A002894 in Sloane's on-line encyclopedia)
can be written as a pullbacked $\, _2F_1$ hypergeometric function:
\begin{eqnarray}
\label{A002894}
\hspace{-0.95in}&&   
(1 \, -16\,x \,+16 \,x^2)^{-1/4} \cdot \, 
_2F_1\Bigl([{{1} \over{12}}, \,{{5} \over{12}} ],\, [1], \, H_4(z)\Bigr)
 \quad  \,
\hbox{with:} \,   \quad  
z \, \, = \, \, {{1 \, -16 \, x} \over {x}}.
\end{eqnarray}
Recalling Maier's paper~\cite{Maier1} one knows that (\ref{A002894}) 
can alternatively 
also be written  a pullbacked $\, _2F_1$ hypergeometric function
with the $\, 12^3/j_4'$ of Table 5 of Maier~\cite{Maier1}. The diagonal 
of $\, 1 / (1\, - x - y - z \, +x\,z)$ is clearly a {\em modular form}. 
As a byproduct this suggests that the diagonal of the rational function of three variables
\begin{eqnarray}
\label{A002894byproduct}
\hspace{-0.95in}&& \quad \quad \quad \quad \quad 
{\frac {z \, \cdot (1 \, -2\,x \, -y) }{ (1 \, -x-y-z \, + x \, z)
 \cdot \,  (1 \, -x -y \, -xz -yz) }}
\end{eqnarray}
is zero. This can be checked directly. 

\vskip .1cm 

\vskip .1cm 

$\bullet$ The diagonal of  $ \, 1/(1 \, - x - y - z \, -x\,y\,z)$, 
which corresponds to the sequence 
$[1, 7, 115, 2371, 54091, 1307377, \, \cdots]$ (oeis number A081798 in 
Sloane's on-line encyclopedia)
can be written as a pullbacked $\, _2F_1$ hypergeometric function:
\begin{eqnarray}
\label{A081798}
\hspace{-0.95in}&&   \quad  \quad \quad
(1\, -27\, x\, +3\, x^2\, - \, x^3)^{-1/4} \cdot \,  (1\, -\, x)^{-1/4} \cdot \, 
_2F_1\Bigl([{{1} \over{12}}, \,{{5} \over{12}} ],\, [1], \, H_3(z)\Bigr)
\nonumber \\
\hspace{-0.95in}&&  \quad  \qquad \quad \quad
 \qquad 
\hbox{with:}   \qquad  \quad \, \, 
  z \, \, = \, \, {{1 \, -30\,x \,+3\,x^2 \,-x^3} \over {x}},
\end{eqnarray}
or more simply as 
\begin{eqnarray}
\label{A081798bis}
\hspace{-0.95in}&&   \qquad  \quad \quad
{{1} \over { 1 \, -x}} \cdot \, 
_2F_1\Bigl([{{1} \over{3}}, \,{{2} \over{3}} ],\, [1], \, 
{{ 27 \, x} \over { (1 \, -x)^3}}\Bigr).
\end{eqnarray}

\vskip .1cm 

$\bullet$ The diagonal of  $ \, 1/(1 \,- x - y - z \, -x\,y \,- x\,z\, - y\,z)$, 
which corresponds to the sequence 
$[1, 12, 366, 13800, 574650, 25335072, \, \cdots]$ (no oeis number)
can be written as a pullbacked $\, _2F_1$ hypergeometric function:
\begin{eqnarray}
\label{nooie}
\hspace{-0.95in}&&  \qquad  \quad \quad
(1\,-48\,x \, -24\, x^2 )^{-1/4} \cdot \, 
_2F_1\Bigl([{{1} \over{12}}, \,{{5} \over{12}} ],\, [1], \, H_3(z)\Bigr)
 \nonumber \\
\hspace{-0.95in}&&  \qquad  \qquad \quad \quad \quad
\qquad 
\hbox{with:}   \qquad   \quad \, \, z \, \, = \, \,
 {{ 1 \,-54\, x \, -27\,x^2} \over {x \cdot \, (x \,+2)}}.
\end{eqnarray}

\vskip .1cm 

$\bullet$ The diagonal of  $ \, 1/(1 \,- x - y \,  - x\, z - y\,z \, + x\,y\,z)$,
 which corresponds to the sequence 
$[1, 3, 25, 243, 2601, 29403, \, \cdots]$ 
(oeis number A245925 in Sloane's on-line encyclopedia)
can be written as a pullbacked $\, _2F_1$ hypergeometric function:
\begin{eqnarray}
\label{A245925}
\hspace{-0.95in}&&   \quad \quad \, \,
(1\,-12 \,x \, -10\,x^2 \, -12\, x^3 \, +x^4)^{-1/4} \cdot \, 
_2F_1\Bigl([{{1} \over{12}}, \,{{5} \over{12}} ],\, [1], \, H_4(z)\Bigr)
\nonumber \\
\hspace{-0.95in}&&  \qquad   \qquad \quad  \quad 
\hbox{with:}   \qquad   \quad  \, \, 
z \, \, = \, \, - \,{{(1\, +x)^2} \over {x}}.
\end{eqnarray}

\vskip .1cm 

$\bullet$ The diagonal of  $ \, 1/(1 \,  - x - y - z \, +x\, y\, z)$, 
which corresponds to the sequence\footnote[2]{Number of 
effective multiple alignments of three equal-length sequences.} 
$[1, 5, 67, 1109, 20251, 391355, \, \cdots]$ (oeis number A124435 in Sloane's 
on-line encyclopedia)
can be written as a pullbacked $\, _2F_1$ hypergeometric function:
\begin{eqnarray}
\label{A124435}
\hspace{-0.95in}&&  \qquad  \quad 
(1\, +x)^{-1/4} \cdot \, (1 \, -21\, x \, +3\, x^2 \, +x^3)^{-1/4} \cdot \, 
_2F_1\Bigl([{{1} \over{12}}, \,{{5} \over{12}} ],\, [1], \, H_3(z)\Bigr)
\nonumber \\
\hspace{-0.95in}&&  \qquad  \qquad  \qquad \quad 
\hbox{with:}   \qquad   \quad  \, \, 
 z \, \, = \, \,
 {{ 1 \,-24\, x \,+3\, x^2 \, + x^3} \over {x}}. 
\end{eqnarray}

\vskip .1cm 

$\bullet$ The diagonal of  $ \, 1/(1 \,  - x + y + z + x\,y + x\,z - y\,z  +x\,y\,z )$, 
which corresponds to the sequence 
\\
$[1, 11, 325, 11711, 465601, 19590491, \, \cdots]$ (no oeis number)
can be written as a pullbacked $\, _2F_1$ hypergeometric function:
\begin{eqnarray}
\label{noeis2}
\hspace{-0.95in}&&    \qquad   \quad 
(1\, \, -46\, x \, +x^2)^{-1/4} \cdot \, (1\, +x)^{-1/2} \cdot \, 
_2F_1\Bigl([{{1} \over{12}}, \,{{5} \over{12}} ],\, [1], \, H_3(z)\Bigr)
\nonumber \\
\hspace{-0.95in}&&  \qquad  \qquad  \qquad  \quad 
 \qquad 
\hbox{with:}   \qquad  \quad  \, \, 
z \, \, = \, \, {{1 \,-52\, x \,  +x^2} \over {2 \, x}}.
\end{eqnarray}

\vskip .2cm 

\section{Two attempts to break the Poincar\'e duality}
\label{attempt}

Along the line, sketched in section (\ref{nonhomo}), which amounts to
 seeking for diagonals of rational functions with 
a ``sufficiently singular'' denominator, let us try to find examples of 
linear differential operators that are {\em not homomorphic to their adjoints}. 
We study here two sets of singular denominators for the rational 
functions, first polynomials 
corresponding to classifications of singular varieties 
performed by V. I. Arnold~\cite{Arnold},
then denominators that factor into two polynomials.  

\vskip .2cm 

\subsection{Diagonals of rational functions associated with singular algebraic varieties}
\label{Arnold}

If one believes that the situations where the linear differential operator
annihilating the diagonal of a rational function should
correspond to situations of  algebraic varieties ``singular enough'' to 
break the Poincar\'e duality~\cite{Griffiths}, 
it is  tempting to study the 
singular algebraic varieties classified by V. I. Arnold~\cite{Arnold}.
The classification of the simplest singularities turned out to be 
related to Lie, Coxeter and Weyl groups, $A_n, \, D_n, \, E_n$,
 and to the classification of platonic
solids in Euclidean three spaces~\cite{Arnold}. 
Arnold's paper gives a set of polynomials of three variables that 
are obvious candidates to be considered as denominators 
of rational functions of three variables:
$\, P \, = \, \, x^2\,z \,+y^3+z^5 \,+2\,y\,z^4,$
$\, x^2\,z+y\,z^2 \,+a \,y^3\,z, \, x^3+y^3+z^4+a\,x\,y\,z^2$,  ... 

With these polynomials, we obtained the linear differential 
operators annihilating the diagonals of the rational functions 
of the form $\, 1/P$. These linear differential operators of various orders 
(order $22$ for $\, Q_{10}$ in~\cite{Arnold}, order $14$
 for  $\, Q_{12}$ in~\cite{Arnold},
order $12$ for $\,S_{11}$, order 8 for $\,S_{12}$, order 10 for $\, U_{12}$, 
order $16$ for 
$\,Z_{13}$, order 18 for $\,W_{12}$, ...) are 
{\em all homomorphic to their adjoints},
their differential Galois groups being 
$\, Sp(n, \, \mathbb{C})$ {\em symplectic} groups, 
the canonical decomposition~\cite{Canonical} being in terms of only 
order-two self-adjoint operators. 

\vskip .2cm 

\subsection{Diagonals of rational functions of six variables}
\label{morex}

Apparently the kind of singular behaviour required to 
``break the Poincar\'e duality'',
considered in V. I. Arnold~\cite{Arnold} for {\em three variables}, 
is not ``sufficiently singular''. It could be that 
these ``sufficiently singular'' situations require much more than 
three variables. Therefore we have considered in this section, 
 rational functions 
of {\em six} variables (see section (\ref{nonhomosub}) 
and (\ref{misnon}) below). Since the theory of singularities of algebraic 
varieties suggests that the situation of ``singular enough'' algebraic 
varieties often correspond to algebraic varieties that {\em factor},
we have studied exhaustively the diagonals of rational functions of the form 
$\, 1/\left( (1 \, -P_1(x, \, y, \, z)) (1 \, -P_2(u, \, v, \, w)) \right)$, 
where the degree of the two polynomials in each of their 
three variables is less than one, and 
their coefficients are $\, 0$ or $\, 1$. Note that 
this also corresponds to considering the Hadamard product
of the  diagonal of  $\, 1/(1 \, -P_1(x, \, y, \, z))$ 
and of the  diagonal of  $\, 1/(1 \, -P_2(u, \, v, \, w))$.
 The number of such classes yielding different diagonals is only $\, 170$. We 
have obtained all the corresponding linear differential operators 
using the Mathematica ``HolonomicFunctions'' package~\cite{Koutschan}.
The order of the (minimal order) linear differential operators runs 
from $\, 2$ to $\, 12$. The number of linear differential operators 
corresponding to the various orders is given in Table (\ref{tab:ODEs}).
\begin{table}[htdp]
\caption{\label{tab:ODEs}
Number of operators corresponding to the various orders, number of 
operators with symplectic and orthogonal differential Galois groups.
 }

\vskip .1cm

\begin{center}
\begin{tabular}{|c|c|c|c|c|c|c|c|c|c|c|c|}\hline
\hline
    Order of Oper.& 2   &  3  &  4 &  5 &  6 &  7  &  8 &  9 &  10 &  11 &  12  \\   \hline
    Number of Oper. & 2   &  3  &  19  &  13 &  39 &  0 &  52  &  0  &  36  &  0 &  6  \\  \hline
    Number of $Sp(n,\mathbb{C})$& 2  &  0 &  7 &  0 & 35 &  0 & 52 &  0  & 36 & 0  &  6  \\  \hline
    Number of $SO(n,\mathbb{C})$& 0  &  3 &  12 &  13 & 4 &  0 &  0 &  0 &  0 &  0  & 0  \\ \hline
\hline
 \end{tabular}
\end{center}
\end{table}

\vskip .1cm

We also give the number of linear differential operators having 
respectively a symplectic and an orthogonal differential Galois group.
Among these $\, 170$ linear differential 
operators, $\, 91$ can be  analyzed using the DEtools command
in order to see that they are homomorphic to their adjoint, find their differential 
Galois group and their canonical decomposition~\cite{Canonical}. The other 
linear differential operators 
are too large to be analyzed that way: they {\em require to switch to a differential 
theta-system}~\cite{Canonical} in order to find the intertwiners to their adjoint, and, 
then the canonical decomposition from a simple euclidean division~\cite{Canonical}.
Note that these calculations ({\em theta-system calculations}) are still quite massive.

\vskip .1cm 

\vskip .1cm 

We found that the three operators 
with a $\, SO(3, \, \mathbb{C})$ 
orthogonal differential Galois group were self-adjoint. 

We found that all the seven order-$4$ linear differential 
operators with a  symplectic
differential Galois group (namely $\, Sp(4,\, \mathbb{C})$) are {\em self-adjoint},
or conjugated to their adjoints (by a simple rational function), except one with a  
$\, (L_2 \cdot \, M_2 \, + \, 1) \cdot \, r(x)$ 
canonical decomposition~\cite{Canonical}. We found that the $\, 12$ 
linear differential operators with 
a $\, SO(4, \, \mathbb{C})$ orthogonal differential Galois group
had a $\, (L_1 \cdot \, M_3 \, + \, 1) \cdot \, r(x)$ canonical 
decomposition~\cite{Canonical}.

All the $13$ order-$5$ linear differential 
operators which have an $\, SO(5,\, \mathbb{C})$ orthogonal 
differential Galois group  have a 
$\, (L_1 \cdot \, M_1 \cdot \,N_3 \,+ \,L_1\,+ \,N_3) \cdot \, r(x) \, $ 
canonical decomposition~\cite{Canonical}.
The four order-$6$ linear differential operators with an $\, SO(6,\, \mathbb{C})$ 
orthogonal differential Galois group have a
$ \, (L_1 \cdot \, M_1 \cdot \, N_1 \cdot \, P_3 \,+ \,L_1 \cdot \, P_3 \,+ \,N_1 \cdot \, P_3  
\,+ \,L_1 \cdot \, M_1 \,+ \, 1) \cdot \, r(x) \, $ canonical 
decomposition~\cite{Canonical}.

We found that the other $\, 35$ order-$6$ linear differential operators 
have a $\, Sp(6, \, \mathbb{C})$  symplectic 
differential Galois group. Among these  $\, 35$  symplectic  operators 
only two have a 
$\, (L_2 \cdot \, M_2 \cdot \, N_2 \, + \,L_2  \, + \,N_4) \cdot \, r(x)$ 
canonical decomposition~\cite{Canonical}, all the other having a 
$\, (L_2 \cdot \, M_4 \, + \, 1) \cdot \, r(x)$ 
canonical decomposition~\cite{Canonical}.

We found that all the $\, 52$ order-$8$ linear differential operators 
have a $\, Sp(8, \, \mathbb{C})$  symplectic 
differential Galois group with a 
$\, (L_2 \cdot \, M_2 \cdot \, N_4 \, + \,L_2  \, + \,N_4) \cdot \, r(x)$ 
canonical decomposition~\cite{Canonical}.

All the $\, 36$ order-$10$  and $\, 6$ order-$12$ 
linear differential operators have respectively $\, Sp(10, \, \mathbb{C})$
and $\, Sp(12, \, \mathbb{C})$ {\em symplectic} differential Galois groups,
with $\, (L_2 \cdot \, M_2 \cdot \, N_2 \cdot \, P_4 \, + \, \cdots) \cdot \, r(x)$ 
and  $\, (L_2 \cdot \, M_2 \cdot \, N_2 \cdot \, P_2 \cdot \, Q_4 \, + \, \cdots) \cdot \, r(x)$
canonical decomposition~\cite{Canonical}. 

\vskip .1cm 

{\bf Remark 1:} In all the decomposition we have obtained,
the rightmost self-adjoint operator~\cite{Canonical}, is always of 
{\em order three} (for orthogonal differential Galois groups)
or of {\em order four} (for symplectic differential Galois groups),
 except an order-$4$ operatoir with a  
$\, (L_2 \cdot \, M_2 \, + \, 1) \cdot \, r(x)$ decomposition and
the two order-$6$ linear differential operators corresponding to 
the diagonals of the two rational functions
\begin{eqnarray}
\label{7}
\hspace{-0.95in}&&  \quad  \quad  \quad 
(1  \, -xy \, -xz \, -yz )  \cdot \, (1 \, -v -w \,  -u v -u w),
 \\
\label{81}
\hspace{-0.95in}&&  \quad  \quad  \quad 
(1  \, -z \, -xy \, -xz \, -yz\, -xyz )  \cdot \, (1 \,-u -v -w \,  -u v w),
\end{eqnarray}
the corresponding operators having a 
$(L_2 \cdot \, M_2 \cdot \,N_2 \,+ \,L_2\,+ \,N_2) \cdot \, r(x)$
canonical decomposition~\cite{Canonical}. 

\vskip .1cm 

{\bf Remark 2:} Using Koutschan's creative telescoping program~\cite{Koutschan}, 
three linear differential operators were quite difficult to obtain
compared to the others. They correspond to the following
denominators of the rational functions:
\begin{eqnarray}
\label{157}
\hspace{-0.95in}&&  \quad  \quad 
(1 \, -x-y-z \, -xz \, -xyz )  \cdot \, (1 \, -u-v-w \,  -uw -vw  \, -uvw),
\\
\label{162}
\hspace{-0.95in}&&  \quad  \quad 
(1 \, -x-y-z \, -xz\, -xyz)  \cdot \, (1 \, -u-v-w \, -uv-uw-vw\, -uvw),
\\
\label{160}
\hspace{-0.95in}&&  \quad  \quad 
(1 \, -x-y-z \, -xz \, -xyz )  \cdot \, (1 \, -u-v-w \, -uv-uw-vw ).
\end{eqnarray}
Koutschan's creative telescoping program gives an order-$12$ linear 
differential operator for (\ref{157}). Note that 
one always needs to verify that the operator obtained 
from this program is actually the {\em minimal order} operator annihilating 
the diagonal\footnote[1]{It may not be irreducible: see, for instance, 
the $\, L_7 \, = \, L_3 \cdot \, L_4$ 
operator associated with the diagonal
 (\ref{diagR}) in section (\ref{orthosymplect}).}. 
Actually, performing {\em Hadamard products}~\cite{CalabiYauIsing,Hadamard} 
modulo primes,  
one can forecast that minimal order for (\ref{157}).
We found an order $\, 10$ which suggests that the order-$12$ linear differential
operator obtained by Koutschan's creative telescoping program~\cite{Koutschan},
is not the {\em minimal order} linear differential 
operator\footnote[2]{One can even forecast the 
degree of the polynomial coefficients 
of that order-$10$ operator: the degree is $\, 74$.}: this order-$12$
{\em actually factorises}. It is the product of {\em two order-one} operators
and of the minimal order-$10$ linear differential operator
annihilating the diagonal: 
$\, L_{12} \, = \, \, L_1 \cdot  \, M_1 \cdot \, L_{10}$. 
The factorization of such very large linear differential
operators\footnote[8]{Using the DFactor command of DEtools in Maple, 
or any other method.} is a {\em quite difficult task}. In a first step 
calculations modulo prime have been performed that enable us to find the 
order of the minimal order operator annihilating all these diagonals. 
We have, however, been able to perform these factorizations
(in characteristic zero, not modulo primes).

\vskip .1cm 

{\bf Remark 3:}  When the linear differential operator annihilating the
diagonal, obtained from Koutschan's program~\cite{Koutschan}, is 
{\em not the minimal order} linear differential operator, one can 
try to obtain that minimal order operator by factorization 
(DFactor in Maple). Unfortunately 
the factorization of (very) large operators like
many of these operators, cannot be obtained using straightforwardly
DFactor DEtools command in Maple. The factorization of the largest operators 
has been obtained, in the following way: one first obtains a large set of
coefficients of the series of the corresponding diagonals (that is the
non-trivial part of the calculation), and then use it as the input of 
a ``guessing procedure\footnote[5]{Essentially of the same type as in gfun.}''.

\vskip .1cm 

{\bf Remark 4:} Koutschan's creative telescoping program~\cite{Koutschan}
 gave us an order-$9$ linear differential operator corresponding to the diagonal of 
$\, 1/((1 \,-x-y-z \,-x\,z \,-x\,y\,z)\,(1 \,-u-v-w \,-u\,w \,-u\,v\,w))$
(i.e. the Hadamard square of the diagonal of 
$\, 1/(1 \,-x-y-z \,-x\,z \,-x\,y\,z)$). In fact this order-$9$ linear 
differential operator
is {\em not minimal}, the minimal order operator being of order $8$.
This program gave us also four order-$11$ linear 
differential operators, like, for instance, the operator corresponding to 
the diagonal of (\ref{162}). Similarly this order-$11$ linear differential operator
is {\em not minimal}, the minimal order operator being of order $10$ (the degree of the
polynomial coefficients is $74$). This is also the case for the three
other, at first sight, order-$11$ linear differential
operators corresponding to the following
denominators of the rational functions:
\begin{eqnarray}
\label{183}
\hspace{-0.95in}&&  \quad  \quad 
(1 \, -z \, -xz \, -yz \ -xy)  \cdot \, (1 \, -u-v-w \,  -uw -vw  \, -uvw),
\\
\label{122}
\hspace{-0.95in}&&  \quad  \quad 
(1 \, -y -z \, -xz\, -xy \, -yz)  \cdot \, (1 \, -u-v-w \, -uv-uw-vw\, -uvw),
\\
\label{154}
\hspace{-0.95in}&&  \quad  \quad 
(1 \, -x -y -z \, -xz\, -xyz)  \cdot \, (1 \, -u-v-w \,-uw-vw ).
\end{eqnarray}

They are {\em not minimal order operators}, the minimal order 
linear differential operators, annihilating the corresponding diagonals 
being of order $10$ (the degree of the
polynomial coefficients being respectively $42, \, 51,  \, 51$). Therefore
one finds that there is {\em no order-$9$ or order-$11$ operators} 
for this set of $170$ diagonals.

\vskip .1cm 

The linear differential operator annihilating the diagonal of (\ref{160}) 
was the most difficult to obtain 
using Koutschan's creative telescoping program\footnote[8]{The program
computes the  ``telescoper'' (the ODE) and the ``certificate''.
It is a general observation that in most examples, the certificate is
much larger than the telescoper~\cite{BostanLairezSalvy}. We just 
need the telescoper. An algorithm
that computes telescopers without computing the corresponding 
certificates, has already been built~\cite{BostanLairezSalvy,Computing}. }. This 
program gives an order-$13$ linear differential operator. From this exact 
differential operator one can obtain as long as necessary series of the diagonal,
and study, modulo some primes, the linear differential operators 
annihilating these series. One finds, that way, that the 
minimal order operator annihilating the diagonal of (\ref{160}) 
is of order $12$ (with degree $85$). We actually factorised this
order-$13$ operator, and obtained the minimal order-$12$ operator
that annihilates this diagonal\footnote[5]{After completion of this work 
we were told by C. Koutschan that giving extra options to ``FindCreativeTelescoping''
enables the program to find directly the minimal order-$12$ operator in 84 hours
CPU time, instead of the 116 hours CPU time we used to obtain the order-$13$ 
operator. With these extra options the telescoper is 232560 bytes when the certificate
requires 32743760 bytes, to be compared with 265432  bytes and 38422496 bytes for 
respectively the telescoper and certificate in the order-$13$ calculation.}.

\vskip .2cm

{\bf Remark 5:} Seeking for the (large ...) linear differential operators
annihilating the diagonals of (\ref{162}), (\ref{160}), it seems
natural to take into account the selected form of all these
diagonals of rational functions of six variables,
which are, actually, the {\em Hadamard product}~\cite{Hadamard,CalabiYauIsing} 
of two diagonals of rational functions of three variables. In the case of  
(\ref{162}), (\ref{160}) (and even (\ref{157})), the linear differential 
operators annihilating these various diagonals of rational functions 
of three variables, are simple order-two operators. One can imagine
to obtain these linear differential operators using the gfun
command ``hadamard product'' of two simple order-two 
operators\footnote[2]{The gfun[hadamardproduct](eq1, eq2, y(z)) command 
determines the linear differential equation satisfied by the Hadamard 
product of two holonomic functions, solutions of the linear
differential equations eq1 and eq2.}. Unfortunately,
the linear differential operators obtained that way are of 
{\em order much larger} than
the one obtained from Koutschan's creative telescoping
of a diagonal of a function of six variables ! The same remarks apply
for the analysis of all the $170$ diagonals of Table (\ref{tab:ODEs}).
 
\vskip .1cm

\vskip .1cm

\vskip .1cm

\vskip .1cm

Let us give, here, miscellaneous, simple examples  
of these diagonal of rational functions of six variables
(all the exhaustive results being given in 
a supplementary material~\cite{supplementary}).
\subsubsection{Order-two operators: pullbacked $\, _2F_1$ hypergeometric function \\}
\label{morexsub}

\vskip .1cm
\vskip .1cm 

\vskip .1cm 

$\bullet$ The diagonal of the rational function of six variables 
\begin{eqnarray}
\label{SS30}
\hspace{-0.95in}&& \, \,   \,  \,  \,                                  
Diag\Bigl( {{1} \over {
 (1 \, -w\, -u\,v\, -u\,v\,w) \cdot \, (1 \,- \,z \, -x \,y)}} \Bigr)
\\
\hspace{-0.95in}&& \quad  \quad  \,    
 \, \,  \,  = \, \, \, \,  \, 
1\, +6\,x\, +78\,{x}^{2}\, +1260\,{x}^{3}\, +22470\,{x}^{4}\,
+424116\,{x}^{5}\, +8305836\,{x}^{6} \, \, + \,\,\cdots 
\nonumber 
\end{eqnarray}
is annihilated by an order-two linear differential operator. This diagonal
is the Hadamard product~\cite{CalabiYauIsing,Hadamard} of the 
diagonal of  $\,1/(1 \,- \,z \, -x \,y)$
and of the diagonal of  $\, 1/(1 \, -w\, -u\,v\, -u\,v\,w)$, namely the {\em two 
simple algebraic functions}:
\begin{eqnarray}
\label{twoalg}
\hspace{-0.95in}&& \quad  \quad  \quad 
(1\, -4\, x)^{-1/2} 
 \\
\hspace{-0.95in}&& \quad \quad  \quad  \,  \quad   \quad  \,  \quad
  \, \,  \,  = \, \, \, \,  \, 
1 \, \, +2 \,x \,+6 \,x^2 \,+20 \,x^3 \,+70 \,x^4
 \,+252 \,x^5 \,+924 \,x^6 \,\,\,+ \, \,\cdots 
\nonumber
 \\
\label{131363}
\hspace{-0.95in}&& \quad  \quad  \quad 
(1\, -6\, x \, +x^2)^{-1/2} 
 \\
\hspace{-0.95in}&& \quad  \, \quad  \quad \quad   \,  \quad   \quad  
  = \, \, \, \, \, 
 1 \,\,\,  +3 \,x\, \,+13\,x^2\,\, +63\,x^3\,\, +321\,x^4
+1683 \,x^5\,+8989 \,x^6 \,\, \,+\,\,\cdots
 \nonumber 
\end{eqnarray}
This diagonal is actually a pullbacked $\, _2F_1$ hypergeometric function: 
\begin{eqnarray}
\label{alternat}
\hspace{-0.95in}&& \quad \,       \quad    
(1 \,-24\,x\,+48\,x^2)^{-1/4} \cdot \,
 \,_2F_1\Bigl([{{1} \over {12}}, \, {{5} \over {12}}], \, [1], \, P_1(x)\Bigr) 
\nonumber   \\
\hspace{-0.95in}&& \quad   \quad  \quad    \quad   \quad    
      \hbox{where:} 
\quad\quad \quad   \quad  
P_1(x)\, \,\, = \, \, \,  \, \, 
6912 \,{\frac {{x}^{4} \cdot \,(1\,-24\,x\, +16\,{x}^{2}) }{
 (1 \,-24\,x\,+48\,x^2)^{3}}}. 
\end{eqnarray}
This {\em Hauptmodul} can be seen to be of the form $\, 12^3/j_2$ 
(see Table 4 of Maier~\cite{Maier1}):
\begin{eqnarray}
\label{alternatJ2}
\hspace{-0.95in}&& \quad 
P_1(x)\, \,\, = \, \, \, \, 12^3 \cdot \, {{z } \over { (z \, +16)^3}} 
 \quad \quad \quad \hbox{with:}  \quad \quad \, \,  z  \,\, = \, \, \,
 {{1 \,-24\,x\,+16\,x^2 } \over { 2\, x^2}}. 
\end{eqnarray}
This diagonal (\ref{SS30}) has another pullbacked 
$\, _2F_1$ hypergeometric function
representation, corresponding to the other pullback 
$\, P_2(x) \, = \, 12^3 \cdot \, z^2/(z\, +256)^3$:
\begin{eqnarray}
\label{P2alternatJ2}
\hspace{-0.95in}&& \quad \,   \,  \quad      
(1 \,-24\,x\,+528\,x^2)^{-1/4} \cdot \,
 \,_2F_1\Bigl([{{1} \over {12}}, \, {{5} \over {12}}], \, [1], \, P_2(x)\Bigr) 
\nonumber \\
\hspace{-0.95in}&& \quad \quad    \quad   \quad  \quad    
        \hbox{where:} 
\quad\quad \quad   \quad  
P_2(x)\, \,\, = \, \, \,  \, \, 
3456 \,{\frac {{x}^{2} \cdot \,(1\,-24\,x\, +16\,{x}^{2})^2 }{
 (1 \,-24\,x\,+528\,x^2)^{3}}}.
\end{eqnarray}
The pullbacked $\, _2F_1$ hypergeometric function (\ref{alternat})
(or equivalently  (\ref{P2alternatJ2})) is thus a {\em modular form}.

Note that this diagonal (\ref{SS30}) is, in fact, actually identical 
to the diagonal of the rational function of three variables
$\, 1 / (1 \, - x - y - z \,- x\,y + x\,z)$.

\vskip .1cm 

$\bullet$  The diagonal of the rational function of six variables 
\begin{eqnarray}
\label{SS41}
\hspace{-0.95in}&& \,  \,   \,                                  
Diag\Bigl( {{1} \over { 
(1 \, -w\, -u\,v\, -u\,v\,w) \cdot \,
 (1 \,- \,z \, -x \,y\, -x \,y\, z)}} \Bigr)
\\
\hspace{-0.95in}&& \quad  \,      
 \, \,  \,  = \, \, \, \,  \, 
1\,\, +9\,x\,+169\,{x}^{2}\,+3969\,{x}^{3}\,+103041\,{x}^{4}\,
+2832489\,{x}^{5}\,+80802121\,{x}^{6}
\,+ \,\,\cdots \nonumber 
\end{eqnarray}
is annihilated by an order-two linear differential operator.
 This diagonal
is the {\em Hadamard square}
 of the diagonal of  $\,1/(1 \,- \,z \, -x \,y\, -x \,y\, z)$, 
namely the Hadamard square of the  algebraic function (\ref{131363}).
 This diagonal
is actually a pullbacked $\, _2F_1$ hypergeometric function: 
\begin{eqnarray}
\label{alternatz}
\hspace{-0.95in}&&     \quad        
(1 \,-36\,x\,+134\,x^2\,-36\,x^3\,+x^4)^{-1/4} \cdot \,
 \,_2F_1\Bigl([{{1} \over {12}}, \, {{5} \over {12}}], \, [1], \, P_1(x)\Bigr) 
\nonumber \\
\hspace{-0.95in}&& \quad   \quad  \, \,   
    \hbox{where:} 
\quad\quad   \, \,  \,       
P_1(x)\, \,\, = \, \, \,  \, \, 
27648 \,{\frac {{x}^{4} \cdot \,(1 \, -x)^2  \cdot \, (1\, -34\, x \, +x^2)}{
 (1 \,-36\,x\,+134\,x^2\,-36\,x^3\,+x^4)^{3}}}.
\end{eqnarray}
It is also be written as  pullbacked $\, _2F_1$ hypergeometric function: 
\begin{eqnarray}
\label{alternatzz}
\hspace{-0.95in}&&  \quad    
(1 \,+444\,x\,+134\,x^2\,+444\,x^3\,+x^4)^{-1/4} \cdot \,
 \,_2F_1\Bigl([{{1} \over {12}}, \, {{5} \over {12}}], \, [1], \, P_2(x)\Bigr) 
\nonumber \\
\hspace{-0.95in}&& \quad \quad  \quad  \, \, 
    \hbox{where:} 
\quad \quad   \, \,  \,     
P_2(x)\, \,\, = \, \, \,  \, \, 
3456 \,{\frac { x \cdot \,(1-x)^2 \cdot \, (1 \,-34\, x \, +x^2)^4 }{
 (1 \,+444\,x\,+134\,x^2\,+444\,x^3\,+x^4)^{3}}}. 
\end{eqnarray}
which shows that this diagonal is a {\em modular form}: 
these two pullbacks can be seen as the Hauptmoduls $\, 12^3/j_4$,
 $\, 12^3/j_4' \, $ in Table 4 and 5 of Maier~\cite{Maier1}:
\begin{eqnarray}
\label{P1}
\hspace{-0.95in}&& \quad    
P_1(x) \, \, = \, \,  \, 
12^3 \cdot \, {{ z \cdot \, (z \, +16)} \over {(z^2 \, +16\,z\, +16)^3 }} 
 \qquad \hbox{where:}  \qquad 
z  \, \, = \, \,  \,{\frac {{x}^{2}-34\,x+1}{2 \, x}}. 
\end{eqnarray}
Note that this diagonal (\ref{SS41}) is, in fact, actually identical 
to the diagonal of the rational function of three variables
$\, 1 / (1 \, - x - y - z \,- x\,y + x\,z - y\,z \, -x\,y\,z)$.

\vskip .1cm

\subsubsection{Order-four and order-eight operators \\}
\label{morexsubmore}

\vskip .1cm

$\bullet$ The diagonal of the rational function of six variables 
\begin{eqnarray}
\label{diagL3}
\hspace{-0.95in}&&  
Diag\Bigl( {{1} \over { 
(1 \, -w\,-u\,v) \cdot \, (1 \, -x \,y \,-x \,z \, -y \,z)}} \Bigr) 
\, \,   = \, \,   \, 
1 \, + 36 \, x^2 \, + 6300 \, x^4 \,   + \, \, \cdots 
\end{eqnarray}
is annihilated by a linear differential operator of order four, $\, L_4$, 
which has an orthogonal differential Galois group $\, SO(4, \, \mathbb{C})$,
with a simple canonical decomposition~\cite{Canonical},  
$\, L_4 \, = \, \, D_x \cdot \, L_3 \, +12$,  where $\, L_3$ 
is an order-three self-adjoint linear differential operator:
\begin{eqnarray}
\label{L3}
\hspace{-0.95in}&&  \quad \quad \quad \quad \, 
L_3 \, \,  = \, \, \, \,  \,   
x^{2} \cdot  \, (432 \,x^{2} \, -1) \cdot \, D_x^3 
\,  \, \,  \, 
+3\, x \cdot \, (864 \,{x}^{2}-1)  \cdot \,  D_x^2 
\nonumber \\
 \hspace{-0.95in}&&  \quad \quad \quad \quad \quad \quad \quad \quad \quad \quad
\, + \, (2868\,{x}^{2} \, -1) \cdot \,  D_x \,\, \,  +276\,x
\end{eqnarray}

$\bullet$ The diagonal of the rational function 
\begin{eqnarray}
\label{diagL4}
\hspace{-0.95in}&& \quad \quad  \quad  \quad 
Diag\Bigl( {{1} \over { (1 \, -u\, v \, -u\, w \, -v\, w)
 \cdot \, (1 \, -x\, y\, -x\, z\, -y\, z)}} \Bigr)
\nonumber \\
\hspace{-0.95in}&&  \quad \quad \quad \quad  \quad \quad  \quad \quad \, 
 \, \,  = \, \, \, \, \,  \,  \, 
1 \, \,\,   + \, 36 \, x^2 \, +\,  8100 \, x^4 \,\,  \,  + \, 2822400 \, x^6
 \,\, \, + \, \, \, \cdots  
\end{eqnarray}
is annihilated by a linear differential operator of order four, $\, L_4$, 
which is self-adjoint, its differential Galois group being 
the symplectic group $\,Sp(4,\mathbb{C})$.

$\bullet$ The diagonal of the rational function 
\begin{eqnarray}
\label{diagL8}
\hspace{-0.95in}&&  \quad   \quad \quad  \, 
Diag\Bigl( {{1} \over { (1 \,-u\,v\,-u\,w\,-v\,w\, -u\,v\,w)
 \cdot \, (1 \, -x\, y\,-x\,z\,-y\,z)}} \Bigr)
\nonumber \\
\hspace{-0.95in}&&  \quad \quad\quad \quad   \quad \quad \, 
 \, \,  = \, \, \,  \, \, \,\, 
1 \, \,  + \,\, 42 \, x^2 \, +  13590 \, x^4 \,\, 
 +7410480 \, x^6 \,   \, + \,  \, \, \cdots  
\end{eqnarray}
is annihilated by a linear differential operator of order eight, $\, L_8$, 
which has a symplectic differential Galois group $\, Sp(8, \, \mathbb{C})$,
with a simple canonical decomposition~\cite{Canonical},  
$L_8 \, = \, \, (L_2 \cdot \,  M_2 \cdot \, N_4 \, + L_2 \, + \,  N_4) \cdot \,r(x)$, 
where $\, L_2$ and $\, M_2$ are two order-two self-adjoint operators and $\, N_4$ is
an {\em order-four self-adjoint} linear differential operator.

\vskip .2cm 

\section{Miscellaneous examples of diagonals of rational functions with
operators that are not homomorphic to their adjoints}
\label{misnon}

\vskip .1cm 

Another slight modification of (\ref{diag3F2a})
 amounts to considering  the diagonal of a rational function 
of {\em six variables}
\begin{eqnarray}
\label{diag3F2d}
\hspace{-0.95in}&&  \, \, \, 
Diag\Bigl(  
{{ 1 \, -9 \, x \, y} \over { (1 - 3\, y - 2\, x + 3\, y^2 + 9\, x^2\, y) 
\cdot \, (1\,-z\,-u\, -u\, z)\cdot \,(1\,-v\,-w)}} 
\Bigr).  
\end{eqnarray}
The corresponding (order-four) linear differential operator is 
{\em not homomorphic to its adjoint}, even with an algebraic extension,
and its differential Galois group is  $\, SL(4, \, \mathbb{C})$.

\vskip .2cm
 
Similarly, the diagonal of the {\em five variable}
 rational function 
\begin{eqnarray}
\label{diag3F2e}
\hspace{-0.95in}&&  \quad \quad \, 
Diag\Bigl(  
{{ 1 } \over { (1\,-x\,+3\,y -27\,x\,y^3\, -27 \,x\,y^2\,-9\,x\,y\,+3\,y^2) 
\cdot \, (1\, - u \,- v\, - u\,z \,- v\,z)}} 
\Bigr),
\nonumber   
\end{eqnarray}
is annihilated by an order-three linear differential operator
which is {\em not homomorphic to its adjoint} (even with 
an algebraic extension), 
and its differential Galois group is  $\, SL(3, \, \mathbb{C})$.

\vskip .1cm 

Similarly, the diagonal of the {\em five variable}
 rational function 
\begin{eqnarray}
\label{diag3F2f}
\hspace{-0.95in}&&  \quad \quad  \quad \quad \, 
Diag\Bigl(  
{{ 1 - 9\, x\, y } \over { (1 - 3\, y - 2\, x + 3\, y^2 + 9\, x^2\, y) 
\cdot \, (1 - u - v  -w)}} 
\Bigr),  
\end{eqnarray}
is annihilated by an order-three linear differential operator
which is {\em not homomorphic to its adjoint},
and its differential Galois group is  $\, SL(3, \, \mathbb{C})$.

\vskip .1cm 

Similarly, the diagonal of the {\em five variable}
 rational function 
\begin{eqnarray}
\label{diag3F2g}
\hspace{-0.95in}&&  \quad \quad \quad \, 
Diag\Bigl(  
{{ 1 - 9\, x\, y } \over { (1 - 3\, y - 2\, x + 3\, y^2 + 8\, x^2\, y) 
\cdot \, (1 - u - v  -w)}} 
\Bigr),   
\end{eqnarray}
is annihilated by an order-five linear differential operator
which is {\em not homomorphic to its adjoint},
and its differential Galois group is  $\, SL(5, \, \mathbb{C})$.

\end{document}